\documentclass[
 reprint, superscriptaddress, amsmath,amssymb, aps, prb, ]{revtex4-2}

\usepackage{CJK}
\usepackage{amsmath}
\usepackage{physics}
\usepackage{amsfonts}
\usepackage{graphicx}
\usepackage[caption=false]{subfig}
\usepackage{placeins}
\usepackage{color}
\usepackage{empheq}
\usepackage{xcolor} 
\usepackage{soul} 

\renewcommand{\vec}[1]{\mathbf{#1}}

\begin{document}

\title{Diffusive Relaxation of Participation Entropy in $\mathrm{U}(1)$-symmetric Dynamics}

\author{Hanchen Liu}
\affiliation{Department of Physics, Boston College, Chestnut Hill, MA 02467, USA}
\author{Tianci Zhou}
\affiliation{Department of Physics, Virginia Tech, Blacksburg, Virginia 24061, USA}
\author{Xiao Chen}
\affiliation{Department of Physics, Boston College, Chestnut Hill, MA 02467, USA}
\date{\today}

\begin{abstract}
Participation entropy (PE) quantifies the spread of a many-body wavefunction across configuration space. While PE relaxes rapidly in generic chaotic systems, we show that $\mathrm{U}(1)$ conservation laws slow it down by imprinting with the slow hydrodynamic modes. Focusing on translationally invariant initial states, we develop a cluster expansion around equilibrium and show that, after local density inhomogeneities decay, the leading PE deficit is dominated by squared connected density correlations. The long time relaxation is therefore controlled by diffusive correlation spreading, giving $\Delta S(t)\sim t^{-1/2}$ in the hydrodynamic regime and crossing over to $\sim \exp[-O(t/L^2)]$ when $t\geq L^2$.  We confirm this entropy correlation relation using exact computation and infinite system tensor network simulations in various quantum $\mathrm{U}(1)$ conserving circuits. Our results 
establish PE as a probe of hydrodynamic memory and suggest that its slow relaxation is a generic consequence of conservation laws.
\end{abstract}
\maketitle

\section{Introduction}
\label{sec:intro}

One fundamental goal in quantum many-body dynamics is to determine the timescales over which initially simple out-of-equilibrium states become complex~\cite{Eisert2015OutOfEquilibrium,DAlessio2016QuantumChaosETH,Rigol2008Thermalization}. Various diagnostics reflect different aspects of this process. Entanglement entropy quantifies basis independent quantum correlations between spatial regions~\cite{Eisert2010AreaLaws} and typically grows ballistically in locally interacting systems~\cite{CalabreseCardy2005EntanglementDynamics,KimHuse2013BallisticEntanglement,Nahum2017EntanglementGrowth}. In contrast, there are other entropy measures which capture how fast a state spreads over a chosen basis. For example, magic (quantified by the stabilizer R\'enyi entropy) measures spreading over Pauli strings, while participation entropy (PE) measures spreading over the computational basis~\cite{Leone2022SRE,Turkeshi2024HilbertSpaceDeloc}. Closely related inverse participation ratios (IPRs) and anticoncentration diagnostics have also been studied in chaotic random tensor-network states and noisy quantum circuits~\cite{Lami2025AnticoncentrationRTN,Sauliere2025NoisyAnticoncentration}. Since these quantities can grow locally throughout the system, they can approach their equilibrium values on much shorter time scales than entanglement~\cite{Dalzell2022Anticoncentration,Turkeshi2024HilbertSpaceDeloc,Turkeshi2025MagicSpreading}. It is thus natural to ask what mechanisms can slow down the relaxation of such basis-resolved measures.

In this work, we study the relaxation of PE in charge-conserving many-body dynamics. The PE is a general entropy measure; given a probability distribution $P_t(\vec{n})$ over a configuration basis $\{\vec{n}\}$ at time $t$, the R\'enyi PEs are defined by
\begin{equation}
    S_{\alpha}(t)=\frac{1}{1-\alpha}
    \log\sum_{\mathbf{n}}P_t(\mathbf{n})^\alpha .
    \label{eq:renyi_pe_definition}
\end{equation}
In the limit $\alpha\to 1$, this expression reduces to the Shannon PE,
\begin{equation}
    S_{1}(t)=-\sum_{\mathbf{n}} P_t(\mathbf{n})\log P_t(\mathbf{n}),
    \label{eq:pe_definition_intro}
\end{equation}

In the quantum context, $P_t(\mathbf{n})$ can be identified with the Born probability of observing the configuration $\vec{n}$. Physically, these entropies capture the effective number of configurations carrying appreciable weight, and thus providing a measure how broadly the state delocalized over the chosen basis.

For generic local unitary dynamics without conservation laws, magic and PE saturate to their equilibrium values rapidly in random-circuit settings~\cite{Fisher2022qey,Turkeshi2024HilbertSpaceDeloc,Turkeshi2025MagicSpreading,Dalzell2022Anticoncentration}. More precisely, their deviations from equilibrium (the entropy deficit) decay exponentially with a system size independent rate, $\sim \exp(-a t)$, which can be inferred from the finite spectral gap~\cite{Turkeshi2024HilbertSpaceDeloc,Turkeshi2025MagicSpreading}. This exponential relaxation toward equilibrium implies a very short $\ln L$ time scale for reaching the equilibrium volume law~\cite{Dalzell2022Anticoncentration,Turkeshi2024HilbertSpaceDeloc,Turkeshi2025MagicSpreading}, which is parametrically shorter than the $O(L)$ time scale for entanglement~\cite{CalabreseCardy2005EntanglementDynamics,KimHuse2013BallisticEntanglement,Nahum2017EntanglementGrowth}. One way to slow down its dynamics is to introduce non-unitarity. In a prior work, we investigated both magic and PE at measurement-induced phase transitions and found that the relaxation exhibits critical slowing down at the critical point~\cite{Heinrich2026MagicPEMIPT}. The deficit still decays exponentially, but with a relaxation time set by $L^z$, where $L$ is the system size and $z$ is the dynamical exponent.

Here we investigate an alternative mechanism for slow PE relaxation: hydrodynamic constraints imposed by a continuous symmetry. Specifically, we focus on dynamics with a conserved $\mathrm{U}(1)$ charge. Charge conservation constrains the evolution of probability distributions in the charge basis and introduces a diffusive relaxation channel. Recent work has also connected conservation laws and transport to the dynamics of PE, coherence, and nonstabilizerness~\cite{Tirrito2025UniversalSpreading,Aditya2026CoherenceConservation}. We show that when the initial local density profile is close to uniform, the local inhomogeneities decay rapidly, but the PE deficit relaxes only diffusively. In one dimension, the corresponding relaxation time scales as $L^2$, and the entropy deficit $\Delta S$ obeys a universal hydrodynamic scaling form:
\begin{equation}
\label{eq:two_part_scaling_function}
\Delta S \sim 
    \begin{cases}
        L t^{-1/2}, & 1\ll t\ll L^2,\\
        \exp[-\gamma t/L^2], & t \gtrsim L^2.
    \end{cases}
\end{equation}

We begin our analysis with a simple classical stochastic process that conserves particle number: the Symmetric Simple Exclusion Process (SSEP), which models interacting particles subject to a hard-core constraint~\cite{kipnis1999scaling,spohn1991large}. Its diffusive dynamics allows us to study PE relaxation in a classical setting. Moreover, its transition rule can also be obtained by averaging over $\mathrm{U}(1)$ charge conserving random circuits~\cite{Rakovszky2018DiffusiveOTOC,Khemani2018OperatorHydro}. At late times, when the probability distribution over configurations enters the long wavelength hydrodynamic regime, expanding the distribution relates the PE deficit to squared connected density correlations. Starting from a translational invariant initial state, the one-body terms decay exponentially. However, the two-body connected correlations relax diffusively due to the conserved charge, which in turn produces the slow diffusive relaxation of PE.

This analysis leads to a broader connection between hydrodynamic modes and the late time relaxation of PE. The probability distribution over configurations approaches the uniform equilibrium distribution only when all the relevant connected correlations vanish. Thus, even though local density fluctuations can decay quickly, the multipoint correlations constrained by charge conservation decay with the diffusive time scale, which determines the PE relaxation. The same mechanism applies more generally whenever the PE deficit can be expanded around the uniform distribution and the relevant connected correlations exhibit diffusive dynamics.

We verify this picture numerically in the SSEP by computing both the connected correlation functions and the PE deficit. We then demonstrate that the same mechanism applies to quantum dynamics with $\mathrm{U}(1)$ symmetry. More specifically, we confirm the predicted scaling in Haar random $\mathrm{U}(1)$-symmetric circuits, where the mapping of the annealed entropy to the SSEP is exact. Finally, we examine more general quantum circuits composed of five-parameter $\mathrm{U}(1)$ symmetric gates. Within the regime where errors are controlled, the observed PE relaxation remains consistent with the proposed scaling in Eq.~\eqref{eq:two_part_scaling_function} once the connected correlation enters into the hydrodynamic regime. 

The rest of the paper is organized as follows. Sec.~\ref{sec:classical} defines the classical SSEP model, introduces the entropy observable, and examines the relaxation of its PE. Sec.~\ref{sec:entropy_corr} derives the entropy expansion that relates the relaxation of non-local correlations to that of PE. Sec.~\ref{sec:quantum_u1} explains how the same mechanism enters Haar random $\mathrm{U}(1)$ symmetric circuits and presents numerical confirmation of the predicted scaling. Sec.~\ref{sec:quantum_spin_chain} uses more structured $\mathrm{U}(1)$-conserving circuits to test the robustness of the relaxation mechanism when microscopic randomness is reduced. Finally, Sec.~\ref{sec:conclusion} summarizes the resulting hydrodynamic picture and discusses possible applications to other basis dependent quantities, including magic.

\section{Classical Toy Model: SSEP}
\label{sec:classical}

It is instructive to first understand PE relaxation in a classical stochastic model, which we believe  is interesting in its own right. We consider the SSEP, where particles obey a hard core constraint but can otherwise hop freely. The total particle number is conserved. This process can also be regarded as the random average of the $\mathrm{U}(1)$ symmetric quantum dynamics discussed in Sec.~\ref{sec:quantum_u1}.

There is at most one particle on each site, and on a chain of length $L$ a configuration is specified by
\begin{equation}
    \mathbf{n}=(n_1,\ldots,n_L),\qquad n_x\in\{0,1\},\qquad x=1,\ldots,L,
\end{equation}
where $n_x=1$ denotes an occupied site and $n_x=0$ denotes an empty site.

\begin{figure}[!htb]
    \centering
    \includegraphics[width=\columnwidth]{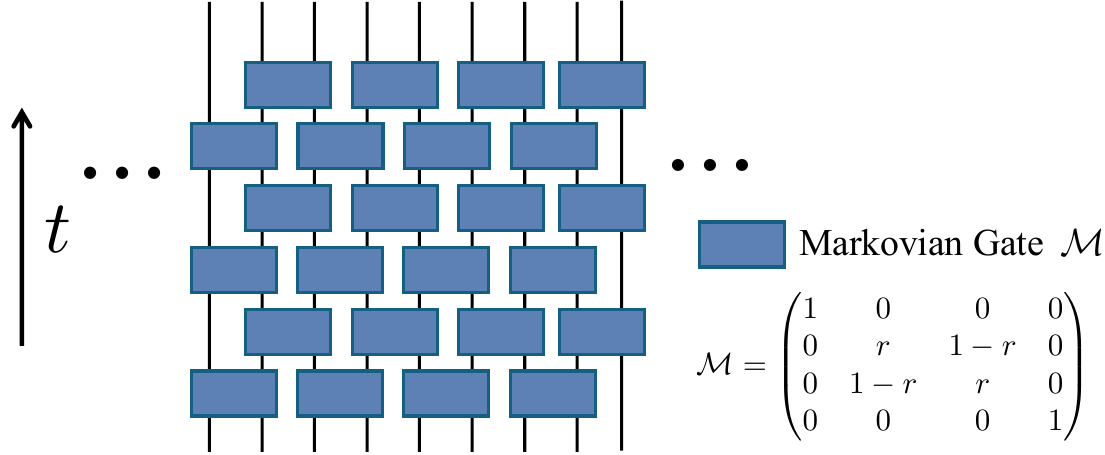}
    \caption{Brickwork update pattern for the one dimensional Markov dynamics.  The two-site stochastic gate is applied first on one sublattice of nearest neighbor bonds and then on the shifted sublattice, giving one full brickwork time step.}
    \label{fig:markov_brickwork}
\end{figure}

We implement the hopping and exclusion of particles locally through a two-site Markov gate. The charge zero and charge two sectors, spanned by $\ket{00}$ and $\ket{11}$, are left invariant. In the one-particle sector $\{\ket{10},\ket{01}\}$, the gate acts as
\begin{equation}
    \mathcal{M}_{Q=1}(r)
    =
    \begin{pmatrix}
        r & 1-r \\
        1-r & r
    \end{pmatrix},
    \label{eq:one_particle_markov_block}
\end{equation}
where columns label the incoming configuration and rows label the outgoing configuration. Thus, with probability $r$ the local configuration is unchanged, while with probability $1-r$ the particle and hole are exchanged, namely, $10 \leftrightarrow 01$. The full chain is evolved in a brickwork circuit by applying this two site Markov gate first to one sublattice of bonds and then to the shifted sublattice, as shown in Fig.~\ref{fig:markov_brickwork}.

The total particle number
\begin{equation}
    N=\sum_{x=1}^L n_x
\end{equation}
is conserved by the dynamics. The number of configurations in the fixed $N$ sector is therefore
\begin{equation}
    \Omega_N=\binom{L}{N}.
    \label{eq:fixed_charge_sector_size}
\end{equation}

For any initial distribution supported in this sector, and for $0<1-r<1$, the dynamics relaxes to the unique stationary distribution, namely the uniform distribution over all configurations with particle number $N$:
\begin{equation}
    P_{\mathrm{eq}}(\mathbf{n})=
    \begin{cases}
        \Omega_N^{-1}, & \sum_x n_x=N,\\
        0, & \text{otherwise}.
    \end{cases}
    \label{eq:ssep_equilibrium_distribution}
\end{equation}

To characterize how the probability distribution approaches its steady state, we monitor the second R\'enyi PE
\begin{equation}
    S_{2}(t)
    =
    -\log \sum_{\mathbf{n}} P_t(\mathbf{n})^2 .
    \label{eq:classical_s2}
\end{equation}
where $P_t( \vec{n} )$ denotes the probability distribution at time $t$.

We focus on its deficit from the steady state value,
\begin{equation}
    \Delta S_2(t)
    =
    \log \Omega_N-S_2(t),
    \qquad
    \delta s_2(t)=\frac{\Delta S_2(t)}{L}.
    \label{eq:classical_s2_deficit}
\end{equation}
Here $\log \Omega_N$ is the second R\'enyi entropy of the uniform distribution in the fixed $N$ sector. Throughout this section, we use the fully mixing gate obtained by setting $r=1/2$ in Eq.~\eqref{eq:one_particle_markov_block},
\begin{equation}
    \mathcal{M}_{Q=1}
    =
    \begin{pmatrix}
        1/2 & 1/2\\
        1/2 & 1/2
    \end{pmatrix}.
    \label{eq:constant_half_markov_gate}
\end{equation}

\begin{figure}[!htbp]
\centering

\includegraphics[width=1\columnwidth]{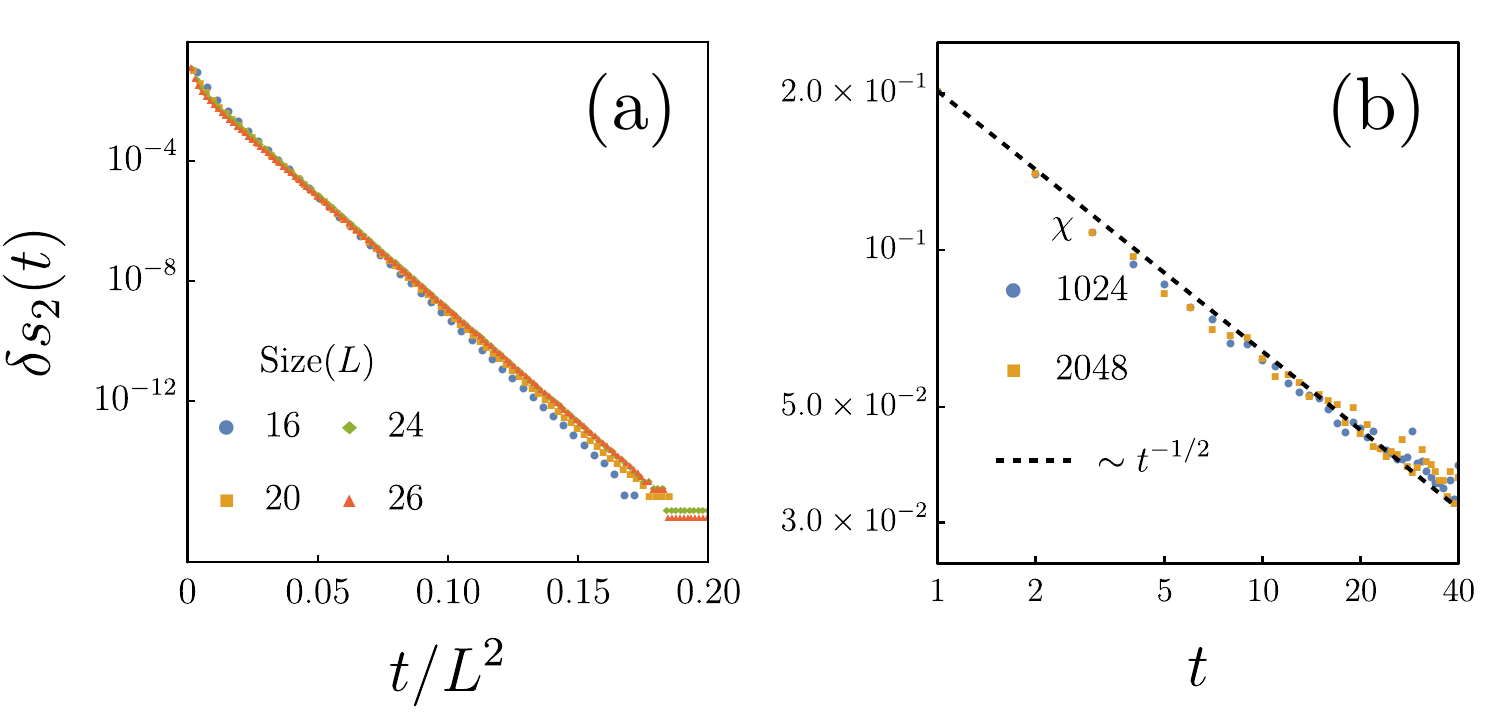}
\caption{Entropy density relaxation in the SSEP brickwork evolution with $r = 1/2$.  Panel (a) shows finite size exact evolution data for the N\'eel initial condition on periodic chains with $L=16, 20, 24, 26$, plotted up to times $t\leq 0.2 L^2$.  Panel (b) shows the corresponding infinite chain calculation obtained using an infinite matrix product state (iMPS) representation with bond dimensions $\chi=1024$ and $2048$.  The dashed guide-to-the-eye indicates the diffusive scaling $t^{-1/2}$.}
\label{fig:markov_entropy_collapse}
\end{figure}

As a simple nonequilibrium initial condition, we take the half filled N\'eel configuration,
\begin{equation}
    P_0(\mathbf{n})
    =
    \delta_{\mathbf{n},\mathbf{n}'},
    \qquad
    \mathbf{n}'=\cdots 01010\cdots ,
\end{equation}
with periodic boundary conditions.

We use two complementary approaches to simulate the brickwork circuit dynamics. To probe times $t\ll L^2$, we exploit translation invariance and evolve the distribution using an infinite matrix product state (iMPS) representation. Operationally, this is an infinite-TEBD style update: two-site gates are applied within a finite unit cell and the resulting tensors are compressed back to a fixed bond dimension~\cite{Vidal2004TEBD,Vidal2007iTEBD,Schollwoeck2011MPSReview,Orus2014TensorNetworksIntro}. Closely related TEBD/MPS methods have been used directly to evolve probability distributions for one-dimensional classical stochastic processes, including exclusion-process dynamics~\cite{JohnsonClarkJaksch2010ClassicalTEBD}. The same tensor-network framework also accommodates nonunitary one-dimensional evolution operators, such as the stochastic Markov gates used here~\cite{OrusVidal2008NonunitaryITEBD,JohnsonClarkJaksch2010ClassicalTEBD}. This allows us to study the early time dynamics directly in the $L\rightarrow\infty$ limit. For the late time regime, $t\gg L^2$, we use exact computation on finite periodic chains. The results are shown in Fig.~\ref{fig:markov_entropy_collapse}. In units where the diffusion constant is absorbed into the time variable, the deficit obeys the following scaling
\begin{equation}
    \delta s_2(t)\sim
    \begin{cases}
        A t^{-1/2}, & 1\ll t\ll L^2,\\
        B\exp[- \gamma t/L^2], & t\gtrsim L^2,
    \end{cases}
    \label{eq:markov_entropy_scaling_summary}
\end{equation}
with nonuniversal constants $A$, $B$, and system size independent constant $\gamma$.

\begin{figure}[t]
    \centering
    \includegraphics[width=\columnwidth]{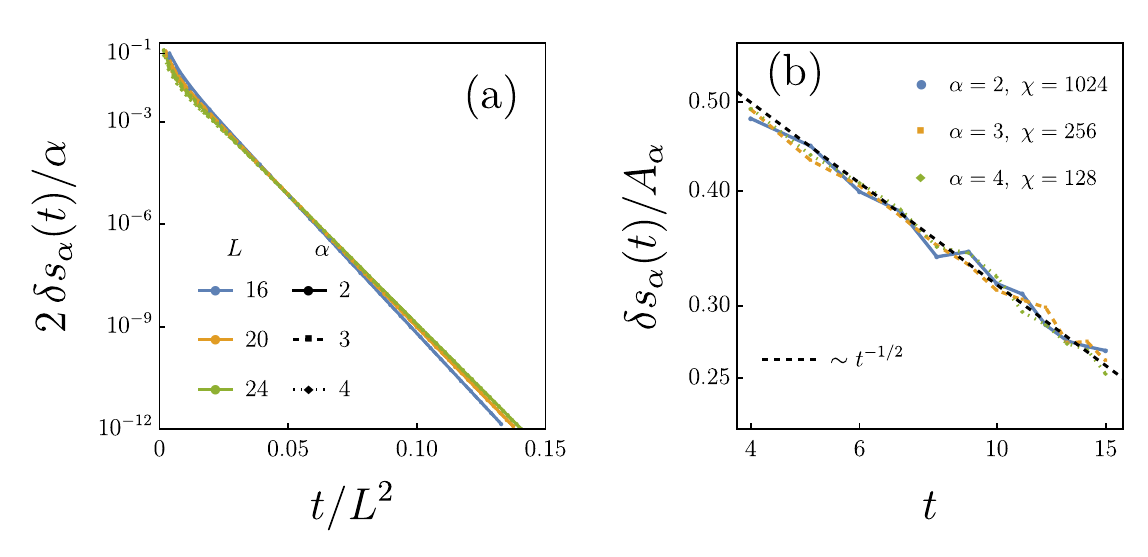}
     \caption{R\'enyi-index dependence of the SSEP entropy deficit. Panel (a) shows periodic-chain exact evolution from the N\'eel state for $\alpha=2,3,4$ and $L=16,20,24$. We plot $(2/\alpha)\delta s_\alpha(t)$ against $t/L^2$. The collapse tests both the finite-size scaling and the leading near-equilibrium prefactor without fitted parameters. Panel (b) shows infinite-chain iMPS data in the common controlled window $4\leq t\leq15$, using bond dimensions $(\chi_2,\chi_3,\chi_4)=(1024,256,128)$ and plotting $\delta s_\alpha(t)/A_\alpha$. The amplitudes $(A_2,A_3,A_4)=(0.195,0.197,0.206)$ are obtained from fixed $t^{-1/2}$ fits over $6\leq t\leq15$; the dashed line denotes $t^{-1/2}$. Later iMPS times are omitted because the higher-R\'enyi entropies become sensitive to bond-dimension truncation.}
    \label{fig:markov_entropy_alpha_collapse}
\end{figure}

\section{Entropy and Correlation}
\label{sec:entropy_corr}

The numerical results in Fig.~\ref{fig:markov_entropy_collapse} demonstrate a slow diffusive relaxation of PE from a N\'eel state. The staggered component of the N\'eel state has high momentum and therefore relaxes on an $O(1)$ time scale. It may thus seem surprising that PE continues to relax slowly. The resolution is that PE measures the relaxation of the full probability distribution toward the global uniform distribution in Eq.~\eqref{eq:ssep_equilibrium_distribution}, while local equilibrium of $P(\vec{n})$ marginalized on a local region can be reached on a much shorter time scale. In the following, we expand the PE deficit around the uniform distribution in Eq.~\eqref{eq:ssep_equilibrium_distribution} in terms of connected correlations. The two-point connected correlations are constrained by the conservation law and relax diffusively. Thus the PE deficit is controlled by the same hydrodynamic modes, giving rise to the diffusive behavior observed in the numerical data.

\subsection{Cluster Expansion}

In the thermalization process, the probability distribution $P_t(\vec{n})$ evolves toward the uniform distribution $P_{\rm eq}(\vec{n})$, and the entropy approaches its maximal value $\ln\Omega_N$~\cite{CoverThomas2006Elements}. We study this convergence by first writing the entropy deficit as a quadratic function of the deviation $\delta P_t(\vec{n})=P_t(\vec{n})-P_{\rm eq}(\vec{n})$, as in the standard relative-entropy expansion around equilibrium~\cite{KullbackLeibler1951Information,yau1991relative}, and then expanding this deviation in cluster density correlations, some of which carry the hydrodynamic modes~\cite{Kubo1962Cumulant,Ruelle1999StatisticalMechanics}.

We consider times when the relative deviation from equilibrium is small, so that in the expansion
\begin{equation}
    P_t(\mathbf{n})
    =P_{\mathrm{eq}}(\mathbf{n})+\delta P_t(\mathbf{n})
    =P_{\mathrm{eq}}(\mathbf{n})\left[1+\epsilon_t(\mathbf{n})\right],
\end{equation}
where we have $\epsilon_t(\vec{n})\ll 1$. Since the entropy is maximized at $P_{\rm eq}(\vec{n})$, the entropy deficit $S_{\rm eq}-S(t)$ is non-negative and starts at quadratic order in $\epsilon_t(\vec{n})$.

We perform the calculation for the R\'enyi PE with fixed index $\alpha>0$, $\alpha\neq 1$,
\begin{equation}
     S_{\alpha}(t)=\frac{1}{1-\alpha}
    \log\sum_{\mathbf{n}}P_t(\mathbf{n})^\alpha .
\end{equation}
At equilibrium, $S_\alpha^{\mathrm{eq}}=\log\Omega_N$ and $P_{\rm eq}(\mathbf n)^\alpha=P_{\rm eq}(\mathbf n)\,e^{(1-\alpha)S_\alpha^{\rm eq}}$. Substituting $P_t(\mathbf{n})=P_{\mathrm{eq}}(1+\epsilon_t)$ gives
\begin{align}
    &\sum_{\mathbf{n}}P_t(\mathbf{n})^\alpha
    =
    e^{(1-\alpha)S_\alpha^{\mathrm{eq}}}
    \sum_{\mathbf{n}} P_{\mathrm{eq}}(\mathbf{n})
    \left(1+\epsilon_t\right)^\alpha
    \nonumber\\
    &=
    e^{(1-\alpha)S_\alpha^{\mathrm{eq}}}
    \left[
        1+
        \frac{\alpha(\alpha-1)}{2}
         \sum_{\mathbf{n}} P_{\mathrm{eq}}(\mathbf{n})\epsilon_t^2
        +O(\epsilon_t^3)
    \right].
    \label{eq:renyi_alpha_ipr_expansion}
\end{align}
The linear term of $\epsilon_t(\vec{n})$ vanishes because $\sum_{\mathbf{n}}\delta P_t(\mathbf{n})=0$. Therefore,
\begin{align}
    \Delta S_\alpha(t)
    &=
    S_\alpha^{\mathrm{eq}}-S_\alpha(t)
    \nonumber\\
    &=
    \frac{1}{\alpha-1}
    \log\left[
        1+
        \frac{\alpha(\alpha-1)}{2}
         \sum_{\mathbf{n}} P_{\mathrm{eq}}(\mathbf{n})\epsilon_t^2
        +O(\epsilon_t^3)
    \right]
    \nonumber\\
    &=
    \frac{\alpha}{2}
    \left\langle\epsilon_t^2\right\rangle_{\mathrm{eq}}
    +O(\epsilon_t^3).
    \label{eq:renyi_alpha_quadratic_distance}
\end{align}
Here, to leading order, the entropy deficit can be written as
\begin{align}
     \left\langle\epsilon_t^2\right\rangle_{\mathrm{eq}}=\sum_{\mathbf{n}}P_{\mathrm{eq}}(\mathbf{n})\epsilon_t(\mathbf{n})^2
     &=
     \sum_{\mathbf{n}}
     \frac{\left[P_t(\mathbf{n})-P_{\mathrm{eq}}(\mathbf{n})\right]^2}
     {P_{\mathrm{eq}}(\mathbf{n})}
     \nonumber\\
     &=
     \Omega_N \sum_{\mathbf{n}} \left[P_t(\mathbf{n})-P_{\mathrm{eq}}(\mathbf{n})\right]^2 .
     \label{eq:weighted_probability_norm}
\end{align}
This is a quadratic distance between the instantaneous distribution and equilibrium, up to a constant prefactor.

The expression $\sum_{\mathbf{n}}P_{\mathrm{eq}}(\mathbf{n})\epsilon_t(\mathbf{n})^2$ averages the relative deviations uniformly over all configurations in the fixed charge sector, and is therefore a measure of the global distance between distributions. This global measure can remain sensitive even when local probes, such as local density fluctuations, cannot distinguish $P_t(\vec{n})$ from $P_{\rm eq}(\vec{n})$.

A cluster expansion makes explicit how this quadratic global distance is related to the multi-point density correlations.

We treat the average over the equilibrium distribution as an inner product for functions on configuration space,
\begin{equation}
\label{eq:eq_measure}
    \langle f( \vec{n} ) , g(\vec{n} ) \rangle_{\rm eq} = \sum_{\vec{n} } P_{\rm eq} (\vec{n} ) f( \vec{n} ) g( \vec{n} ) = \langle f(\vec{n} ) g(\vec{n} ) \rangle_{\rm eq}. 
\end{equation}
Then we introduce local density fluctuations
\begin{equation}
    \delta n_x=n_x-\bar{\rho},
    \qquad
    \kappa=\langle (\delta n_x)^2\rangle_{\mathrm{eq}}
    =\bar{\rho}(1-\bar{\rho}).
\end{equation}
where $\bar{\rho}= \langle n_x \rangle_{\rm eq} $.

Ignoring corrections from the global number constraint, products of distinct $\delta n_x$ form a complete orthogonal basis under the local equilibrium measure in Eq.~\eqref{eq:eq_measure}. We can therefore expand the relative deviation as
\begin{equation}
    \epsilon_t(\mathbf{n})
    =
    \sum_x C_x^{(1)}(t)\delta n_x
    +
    \sum_{x<y} C_{xy}^{(2)}(t)\delta n_x\delta n_y
    +\cdots .
    \label{eq:cluster_expansion}
\end{equation}
The expansion coefficients can be computed using this inner product, as shown in Appendix~\ref{app:cluster_projection}. The first few terms are
\begin{align}
    C_x^{(1)}(t)
    &=
    \frac{\langle \delta n_x\rangle_t}{\kappa},\\
    C_{xy}^{(2)}(t)
    &=
    \frac{\langle \delta n_x\delta n_y\rangle_t}{\kappa^2}.
\end{align}
Here $\langle \cdots\rangle_t$ denotes an average over the instantaneous distribution $P_t(\vec{n})$.

At late times, the one-point density has already relaxed, meaning that $\bar{\rho}-\langle n_x\rangle_t$ is exponentially small. Therefore the second coefficient $\langle \delta n_x\delta n_y\rangle_t$ is equal, up to exponentially small one body corrections, to the connected pair correlator
\begin{equation}
    C(x,y;t)
    =
    \langle n_x n_y\rangle_t
    -
    \langle n_x\rangle_t\langle n_y\rangle_t .
\end{equation}
Substituting Eq.~\eqref{eq:cluster_expansion} into the R\'enyi-$\alpha$ quadratic form in Eq.~\eqref{eq:renyi_alpha_quadratic_distance} and using orthogonality gives a Parseval type result:
\begin{align}
 \Delta S_\alpha(t)
    &=
    \frac{\alpha}{2}
    \bigg[
    \frac{1}{\kappa}
    \sum_x \langle \delta n_x\rangle_t^2
    \nonumber\\
    &\quad
    +
    \frac{1}{\kappa^2}
    \sum_{x<y} C(x,y;t)^2
    +\cdots
    \bigg]
    +O(\epsilon_t^3).
    \label{eq:entropy_correlation_decomposition}
\end{align}
The omitted terms contain higher connected clusters.

The cluster expansion shows that the deficit of the R\'enyi PE of index $\alpha$ has the same leading order time dependence as the second R\'enyi PE, up to an $\alpha$-dependent prefactor. This leading order approximation is valid in the regime where $\epsilon_t(\vec{n})$ is small. We test the dependence on both $\alpha$ and time in Fig.~\ref{fig:markov_entropy_alpha_collapse}. Fig.~\ref{fig:markov_entropy_alpha_collapse}~(a) shows that the relation $\delta s_\alpha\simeq(\alpha/2)\delta s_2$ yields an $\alpha$-independent exponential decay at late times as a function of $t/ L^2$, consistent with Eq.~\eqref{eq:renyi_alpha_quadratic_distance}. In the early time regime $1 \ll t \ll L^2$, Fig.~\ref{fig:markov_entropy_alpha_collapse}~(b) demonstrates a $t^{-1/2}$ decay of the R\'enyi deficit, with a weak $\alpha$-dependence. We explain this weak $\alpha$-dependence in Sec.~\ref{subsec:L_infty} by considering the neglected higher order terms in Eq.~\ref{eq:entropy_correlation_decomposition}.

\subsection{Inhomogeneous initial states}

When the initial density profile is not translationally invariant, the one-point function in Eq.~\eqref{eq:entropy_correlation_decomposition}
provides the leading source of diffusive PE relaxation.

Consider a product configuration chosen randomly from the sector with fixed particle number \(N=\bar\rho L\).  For each realization,
\begin{equation}
    \delta\rho_x(0)=\langle n_x\rangle_0-\bar\rho
\end{equation}
contains fluctuations on all spatial scales, while the fixed-charge
constraint removes the zero-momentum component.  At long wavelengths,
the Fourier modes,
\begin{equation}
    \delta\rho_k(t)\simeq e^{-Dk^2t}\delta\rho_k(0),
\end{equation}
relax diffusively.  The one-body contribution in Eq.~(25) therefore
gives
\begin{equation}
    \langle{\delta s_{\alpha}(t)}\rangle_{\rm ic}\simeq
\frac{\alpha}{2\kappa L}
\sum_{k\neq 0}
e^{-2Dk^2t}
\langle{|\delta\rho_k(0)|^2}\rangle_{\rm ic},
\end{equation}
where $\delta\rho_k=L^{-1/2}\sum_x e^{-ikx}\delta\rho_x$, and $\langle \ldots \rangle_{\rm ic}$ denotes the averaging over initial condition.  For random fixed-filling configurations,
\(\langle{|\delta\rho_k(0)|^2}\rangle_{\rm ic}\simeq\kappa\) for
\(k\neq0\).  Consequently,
\begin{equation}
    \langle{\delta s_{\alpha}(t)}\rangle_{\rm ic}
\sim
\begin{cases}
A_\alpha t^{-1/2}, & 1\ll t\ll L^2,\\[2mm]
B_\alpha(L)\exp[-\gamma_1 t/L^2], & t\gtrsim L^2.
\end{cases}
\end{equation}
Thus, for spatially inhomogeneous initial states, the one-body
density sector alone produces diffusive PE relaxation.  By contrast,
for an exactly translationally invariant initial state, this contribution is absent or decays exponentially fast, and the slow algebraic relaxation is instead controlled by
the connected two-point term in Eq.~(25). This is the main focus of this paper. We will analyze this in the next subsection.

\subsection{Translational invariant initial states}
Consider the connected density correlator
\begin{equation}
    C(x,y;t)
    =
    \langle n_x n_y\rangle_t
    -
    \langle n_x\rangle_t\langle n_y\rangle_t .
\end{equation}
Particle number conservation implies the exact sum rule
\begin{align}
    \sum_y C(x,y;t)
    &=
    \left\langle n_x \sum_y n_y\right\rangle_t
    -
    \langle n_x\rangle_t
    \left\langle \sum_y n_y\right\rangle_t
    \nonumber\\
    &=
    \langle n_x N\rangle_t-\langle n_x\rangle_t\langle N\rangle_t
    =
    0,
    \label{eq:correlation_sum_rule_fixed_n}
\end{align}
where we used the fixed charge constraint $\sum_x n_x=N$.

For a translationally invariant state, or after averaging the origin over one iMPS unit cell, we may write the correlator as a function of the relative coordinate $r=y-x$. In the latter case, we use $C(r,t)$ as shorthand for $L_{\mathrm{cell}}^{-1}\sum_{x=1}^{L_{\mathrm{cell}}}C(x,x+r;t)$. The sum rule then becomes
\begin{equation}
    \sum_r C(r,t)=0,
    \qquad
    \sum_{r\neq 0}C(r,t)=-C(0,t).
    \label{eq:correlation_sum_rule_relative}
\end{equation}
Thus the onsite variance $C(0,t)$ must be balanced by an opposite sign offsite correlation around it. Starting from the N\'eel state, the onsite term relaxes rapidly to its local equilibrium value, while the compensating offsite correlations can only spread through local charge conserving dynamics. We therefore expect the offsite correlator, for $r\neq0$, to obey an effective diffusion equation at long wavelengths,
\begin{equation}
    \partial_t C(r,t)
    \simeq
    D\,\partial_r^2 C(r,t),
    \label{eq:relative_correlation_diffusion}
\end{equation}
with the total offsite weight fixed by Eq.~\eqref{eq:correlation_sum_rule_relative}. In the units used below, we absorb the diffusion constant $D$ into the definition of time.

Since the offsite weight is approximately fixed by $-C(0,t)$ after the fast local transient, it is an $O(1)$ quantity. This initially short ranged weight then spreads over a diffusive length scale
\begin{equation}
    \ell(t)\sim t^{1/2}.
\end{equation}
The typical magnitude of the offsite correlator is therefore of order $\ell(t)^{-1}\sim t^{-1/2}$. Consequently,
\begin{equation}
    \sum_r C(r,t)^2
    \sim
    \ell(t)\,[\ell(t)^{-1}]^2
    \sim
    t^{-1/2}.
    \label{eq:corr_square_diffusive_estimate}
\end{equation}
Through Eq.~\eqref{eq:entropy_correlation_decomposition}, this gives the early time scaling of the entropy deficit.

\begin{figure}[!htbp]
    \centering

    \includegraphics[width=1\columnwidth]{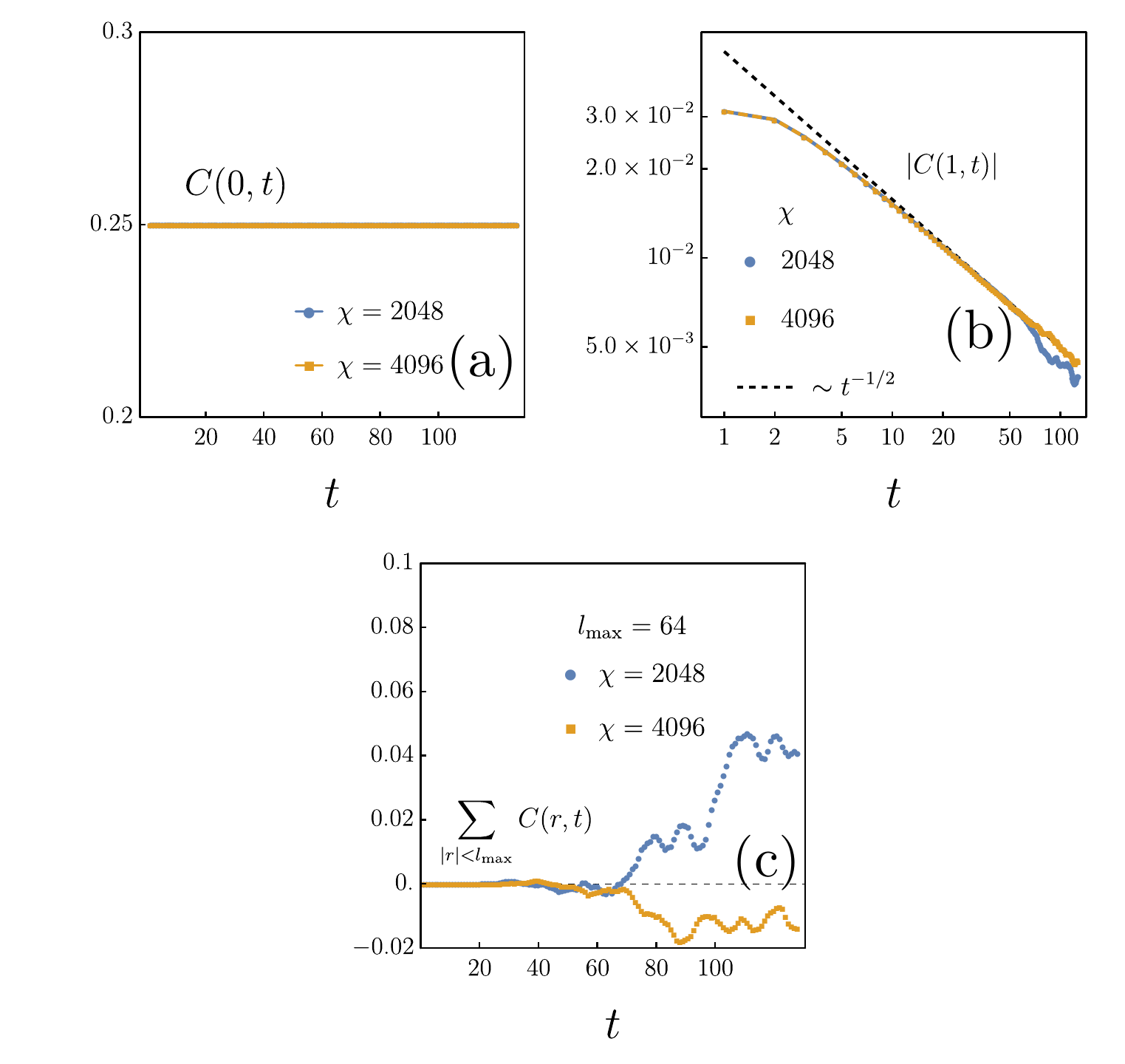}
    
    \caption{Local and conservation diagnostics for the infinite chain Markov iMPS calculation, with bond dimension $\chi=2048, 4096$.  Panel (a) shows that the onsite variance $C(0,t)$ quickly reaches the half filled steady state value $\kappa=1/4$, providing an $O(1)$ local source for offsite correlations.  Panel (b) shows the nearest neighbor absolute connected correlation $|C(1,t)|$ with a $t^{-1/2}$ guide.  Panel (c) checks the finite window version of the conservation sum rule $\sum_{|r|<l_{\rm max}} C(r,t)$ as a function of time. We check that at $t < l_{\rm max}$, $\sum_{|r|<l_{\rm max}} C(r,t)=0$, consistent with our expectation, indicating that the data in (a) and (b) with $t < l_{\rm max}$ has controlled errors.   }
    \label{fig:markov_corr_local_diagnostics}
\end{figure}

\begin{figure}[!htbp]
    \centering
    \includegraphics[width=\columnwidth]{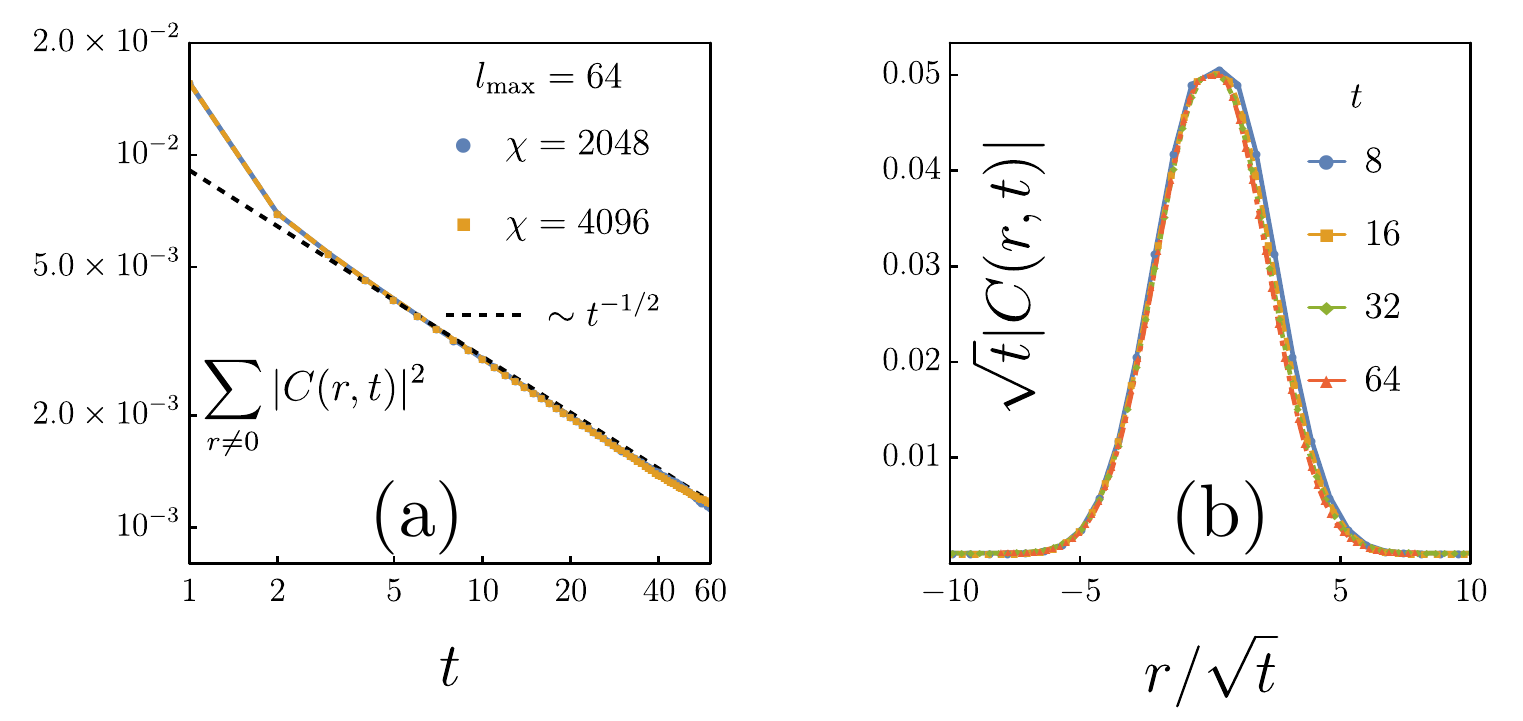}
    \caption{Diffusive scaling diagnostics for the $r = 1/2$ SSEP brickwork evolution with the iMPS method.  Panel (a) shows the algebraic decay of the offsite squared correlation sum, with a $t^{-1/2}$ guide.  Panel (b) shows the diffusive collapse of the offsite ($r \neq 0$) correlation profile under the scaling variable $r/\sqrt{t}$.}
    \label{fig:markov_corr_diffusive_scaling}
\end{figure}

Figs.~\ref{fig:markov_corr_local_diagnostics} and \ref{fig:markov_corr_diffusive_scaling} provide numerical checks of this picture using the iMPS calculation at half filling. The onsite term $C(0,t)$ rapidly approaches its local equilibrium value $\kappa=1/4$, confirming that the local part of the N\'eel pattern melts on a short time scale. The nearest neighbor correlator follows $C(1,t)\sim 1/\sqrt{t}$. Meanwhile, the finite window sum of the connected correlator is an error indicator of the algorithm; it remains close to zero when $t<l_{\rm max}$, consistent with the conservation law constraint in Eq.~\eqref{eq:correlation_sum_rule_relative} (See Fig.~\ref{fig:markov_corr_local_diagnostics} (c)). The offsite squared sum decays with the same $t^{-1/2}$ scaling that controls the entropy deficit in Eq.~\eqref{eq:entropy_correlation_decomposition}.

The full spatial profile gives a more direct test of the diffusive interpretation. At intermediate times before finite size effects become important, the offsite correlator takes the scaling form
\begin{equation}
    C(r,t)
    \simeq
    -t^{-1/2} f\!\left(\frac{r}{\sqrt{t}}\right),
    \label{eq:diffusive_correlation_scaling_form}
\end{equation}
up to nonuniversal normalization factors. Equivalently, $t\,C(r,t)^2$ is controlled by the same scaling variable $r/\sqrt{t}$. Integrating over $r$ then reproduces Eq.~\eqref{eq:corr_square_diffusive_estimate}.

The same correlation based argument also explains the late time finite size crossover in Eq.~\eqref{eq:markov_entropy_scaling_summary}. For $t\gtrsim L^2$, the offsite correlation has spread across the entire periodic chain, and the relaxation is controlled by the lowest nonzero diffusive mode. Fourier mode with momentum $k$ in diffusion equation decays as $\exp( - D k^2 )$. With the smallest momentum $k \sim \frac{1}{L}$, the connected correlator's decay is dominated by 
\begin{equation}
    C(r,t)\sim f_L(r)\exp\!\left[-\frac{\gamma t}{2L^2}\right],
\end{equation}
where $f_L(r)$ is the corresponding spatial mode on the periodic chain. Since the entropy deficit is controlled by the square of the connected correlator, summing over space gives
\begin{equation}
    \Delta S_2(t)\sim \exp\!\left[-\frac{\gamma t}{L^2}\right].
\end{equation}
This yields the late time relaxation form quoted in Eq.~\eqref{eq:markov_entropy_scaling_summary}. A more detailed derivation is given in Appendix~\ref{app: crossover}.

\subsection{Limit of $L\rightarrow\infty$}
\label{subsec:L_infty}

In the regime $1\ll t\ll L^2$, the time dependence is consistent
across different R\'enyi entropies, but the prefactor does not exactly
match $\alpha/2$.  We attribute this to the fact that the deviation
$\epsilon_t(\mathbf n)$ is no longer uniformly small, so the expansion
of the entropy deficit cannot be truncated at the leading order
$O(\epsilon_t^2)$.

To compare with the numerical results, we consider the leading-time
contributions from all powers $\epsilon_t^m$.  For integer
$\alpha\geq2$, normalization gives
$\langle\epsilon_t\rangle_{\rm eq}=0$, and the exact finite expansion is
\begin{equation}
    e^{(\alpha-1)S_\alpha^{\rm eq}}
    \sum_{\mathbf n}P_t(\mathbf n)^\alpha
    =
    1+
    \sum_{m=2}^{\alpha}
    \binom{\alpha}{m}
    \langle\epsilon_t^m\rangle_{\rm eq}.
    \label{eq:renyi_exact_moment_identity}
\end{equation}
Therefore, the exact entropy deficit is
\begin{equation}
    \Delta S_\alpha(t)
    =
    \frac{1}{\alpha-1}
    \log\!\left[
    1+
    \sum_{m=2}^{\alpha}
    \binom{\alpha}{m}
    \langle\epsilon_t^m\rangle_{\rm eq}
    \right].
    \label{eq:renyi_exact_deficit_all_moments}
\end{equation}
For the leading-time power counting, we expand the logarithm to first
order,
\begin{equation}
\begin{aligned}
    \Delta S_\alpha(t)
    ={}&
    \frac{1}{\alpha-1}
    \sum_{m=2}^{\alpha}
    \binom{\alpha}{m}
    \langle\epsilon_t^m\rangle_{\rm eq}\\
    &+
    O\!\left[
    \left(
    \sum_{m=2}^{\alpha}
    \binom{\alpha}{m}
    \langle\epsilon_t^m\rangle_{\rm eq}
    \right)^2
    \right].
\end{aligned}
\label{eq:renyi_finite_moment_expansion}
\end{equation}
We will show that the connected ring in each
$\langle\epsilon_t^m\rangle_{\rm eq}$ scales as $t^{-1/2}$.
Consequently, products of rings decay at least as $t^{-1}$ and do not
contribute to the leading time dependence.

We now compute the contributions of order $t^{-1/2}$.  The one-point
coefficient in the cluster expansion decays exponentially for translational invariant systems, while the
higher connected coefficients are subleading.  We
therefore retain the two-point term,
\begin{equation}
    \epsilon_t(\mathbf n)
    \sim
    \sum_{x<y}
    C_{xy}^{(2)}(t)\delta n_x\delta n_y.
\end{equation}
In the $m$-th power of $\epsilon_t$, one connected contraction gives an
$m$-ring.  Using
$\langle\delta n_x\delta n_y\rangle_{\rm eq}=\kappa\delta_{xy}$, its
contribution is
\begin{equation}
\begin{aligned}
    \left.\langle\epsilon_t^m\rangle_{\rm eq}\right|_{\rm ring}
    \sim{}&
    \kappa^m
    \int_{\mathbb R^m}dx_1\cdots dx_m\,
    \prod_{i=1}^{m}C_{x_ix_{i+1}}^{(2)}(t),
\end{aligned}
\label{eq:finite_m_ring}
\end{equation}
where $x_{m+1}=x_1$.  Using
$C_{xy}^{(2)}(t)=C(x,y;t)/\kappa^2$ and translation invariance,
$C(x,y;t)=C(r,t)$.  Changing to $r_i=x_i-x_{i+1}$ and one
center-of-mass coordinate, whose integration gives $L$, we obtain
\begin{equation}
\begin{aligned}
    \left.\langle\epsilon_t^m\rangle_{\rm eq}\right|_{\rm ring}
    \sim{}&
    \frac{L}{\kappa^m}
    \int dr_1\cdots dr_m\,
    \prod_{i=1}^{m}C(r_i,t)\,
    \delta\!\left(\sum_{i=1}^{m}r_i\right).
\end{aligned}
\end{equation}
Substituting Eq.~\eqref{eq:diffusive_correlation_scaling_form} and
setting $r_i=\sqrt t\,u_i$ gives
\begin{equation}
\begin{aligned}
    \left.\langle\epsilon_t^m\rangle_{\rm eq}\right|_{\rm ring}
    \sim{}&
    \frac{(-1)^mL}{\kappa^m\sqrt t}
    \int du_1\cdots du_m
    \\[-1mm]
    &\quad\times
    \prod_{i=1}^{m}f(u_i)\,
    \delta\!\left(\sum_{i=1}^{m}u_i\right).
\end{aligned}
\label{eq:finite_m_ring_scaling}
\end{equation}
For every fixed $m$, the remaining integral is independent of time.  If
$f$ is approximated by a normalized Gaussian, it is proportional to
$1/\sqrt m$.  Suppressing time-independent normalization and
contraction factors, we therefore have
\begin{equation}
    \left.\langle\epsilon_t^m\rangle_{\rm eq}\right|_{\rm ring}
    \sim
    (-1)^m\frac{L}{\sqrt{mt}}.
    \label{eq:one-ring}
\end{equation}
Terms in $\langle\epsilon_t^m\rangle_{\rm eq}$ decays with $t^{-\text{number of rings}/2}$. Therefore the one-ring term in Eq.~\ref{eq:one-ring} is the leading contribution. 

Collecting the finite set of ring contributions gives
\begin{equation}
    \Delta S_\alpha(t)
    \sim
    \frac{L}{(\alpha-1)\sqrt t}
    \sum_{m=2}^{\alpha}
    (-1)^m\binom{\alpha}{m}\frac{1}{\sqrt m}.
    \label{eq:renyi_finite_ring_sum}
\end{equation}
Within this estimate, the predicted amplitude ratio is
$\Delta S_2(t):\Delta S_3(t):\Delta S_4(t)\simeq1:1.09:1.15$.
For direct comparison, the fitted amplitudes in
Fig.~\ref{fig:markov_entropy_alpha_collapse}(b) are
$(A_2,A_3,A_4)=(0.19544,0.19729,0.20640)$, which give
$A_2:A_3:A_4\simeq1:1.009:1.056$. Thus the one-ring estimate explains why the prefactor deviates from $\frac{\alpha}{2}$, and instead shows a weak increase with $\alpha$. It shows that cluster expansion can capture the scalings of the PE relaxation accurately up to the prefactor in the regime of $1 \ll t \ll L^2$.

\section{Quantum Haar random circuit with $\mathrm{U}(1)$ symmetry}
\label{sec:quantum_u1}

  We now turn to the charge conserving Haar random circuit~\cite{Rakovszky2018DiffusiveOTOC,Khemani2018OperatorHydro,Hearth2025NumberConservingDesigns}.
  The system is a spin $1/2$ chain written in the computational, or charge, basis,
\begin{equation}
    \ket{\mathbf{n}}=\ket{n_1,\ldots,n_L},
    \qquad
    n_x\in\{0,1\},
    \qquad
    x=1,\ldots,L .
\end{equation}
The total charge
\begin{equation}
    N=\sum_x n_x
\end{equation}
is conserved by the dynamics.
  Each nearest neighbor gate conserves the two-site charge and is therefore block diagonal in the basis $\{\ket{00},\ket{01},\ket{10},\ket{11}\}$:
\begin{equation}
    U=
    e^{\mathrm{i}\theta_0}\oplus V_{Q=1}\oplus e^{\mathrm{i}\theta_2},
    \qquad
    V_{Q=1}\in U(2),
\end{equation}
where $V_{Q=1}$ is drawn independently from the Haar measure.

The connection with the classical Markov dynamics is immediate at the level of annealed diagonal probabilities, as in standard treatments of charge-conserving random circuits~\cite{Fisher2022qey,Rakovszky2018DiffusiveOTOC,Khemani2018OperatorHydro}. For a single two-site gate, the $Q=0$ and $Q=2$ sectors are left invariant. In the one-particle sector, Haar invariance and normalization give uniform transition probabilities,
\begin{align}
    \overline{W}_{01\to 01}
    =
    \overline{W}_{01\to 10}
    =
    \overline{W}_{10\to 01}
    =
    \overline{W}_{10\to 10}
    =
    \frac{1}{2}.
\end{align}
Thus the circuit averaged diagonal probability distribution evolves under the same local stochastic rule as Eq.~\eqref{eq:constant_half_markov_gate}, up to the diffusion constant set by the brickwork convention. More explicitly, for
\begin{equation}
    P_t^\psi(\mathbf{n})
    =
    \left|\braket{\mathbf{n}}{\psi(t)}\right|^2 ,
\end{equation}
the annealed distribution $\overline{P_t^\psi(\mathbf{n})}$ follows the same SSEP type evolution as the classical model discussed above. This mapping explains why the hydrodynamic scaling found in the classical dynamics also controls the slow relaxation of PE in the quantum circuit, in close analogy with hydrodynamic tails in other R\'enyi-type observables~\cite{Rakovszky2019SubballisticRenyi}.

We next test this expectation numerically for the quenched averaged PE in Haar random charge conserving circuits.  As in the classical case, we use two complementary methods: infinite chain MPS, implemented by the same finite-unit-cell iMPS/TEBD update~\cite{Vidal2004TEBD,Vidal2007iTEBD,Schollwoeck2011MPSReview}, for $t \ll L^2$ and finite size exact evolution for $t \gg L^2$. For each circuit realization, we evolve a quantum trajectory and compute the second R\'enyi PE
\begin{equation}
    S_2^\psi(t)
    =
    -\log \sum_{\mathbf{n}}\left[P_t^\psi(\mathbf{n})\right]^2 ,
    \label{eq:s2_pe}
\end{equation}
where the superscript $\psi$ emphasizes that this is a trajectory level or quenched quantity. We then average this entropy over circuit realizations.

The steady state value of this quenched quantity is not the classical value $\log\Omega_N$. Instead, a Haar random state in a Hilbert space of dimension
\begin{equation}
    \mathcal{D}=\Omega_N
\end{equation}
has Porter-Thomas fluctuations in its computational basis amplitudes~\cite{PorterThomas1956}. These fluctuations enhance the collision probability relative to a perfectly flat distribution. As derived in Appendix~\ref{app:pt_plateau}, the corresponding second R\'enyi plateau is
\begin{equation}
    S_{2,\mathrm{Haar}}^{\mathrm{eq}}
    \simeq
    \log\frac{\mathcal{D}+1}{2}.
    \label{eq:s2_haar_plateau}
\end{equation}
We therefore define the positive entropy deficit density as
\begin{equation}
    \overline{\Delta S_2^\psi(t)}
    =
    S_{2,\mathrm{Haar}}^{\mathrm{eq}}
    -
    \overline{S_2^\psi(t)},
    \qquad
    \overline{\delta s_2(t)}
    =
    \frac{
        \overline{\Delta S_2^\psi(t)}
    }{L},
    \label{eq:quantum_entropy_deficit}
\end{equation}
where the overline denotes the circuit average.

\begin{figure}[!htbp]
\centering
\includegraphics[width=\columnwidth]{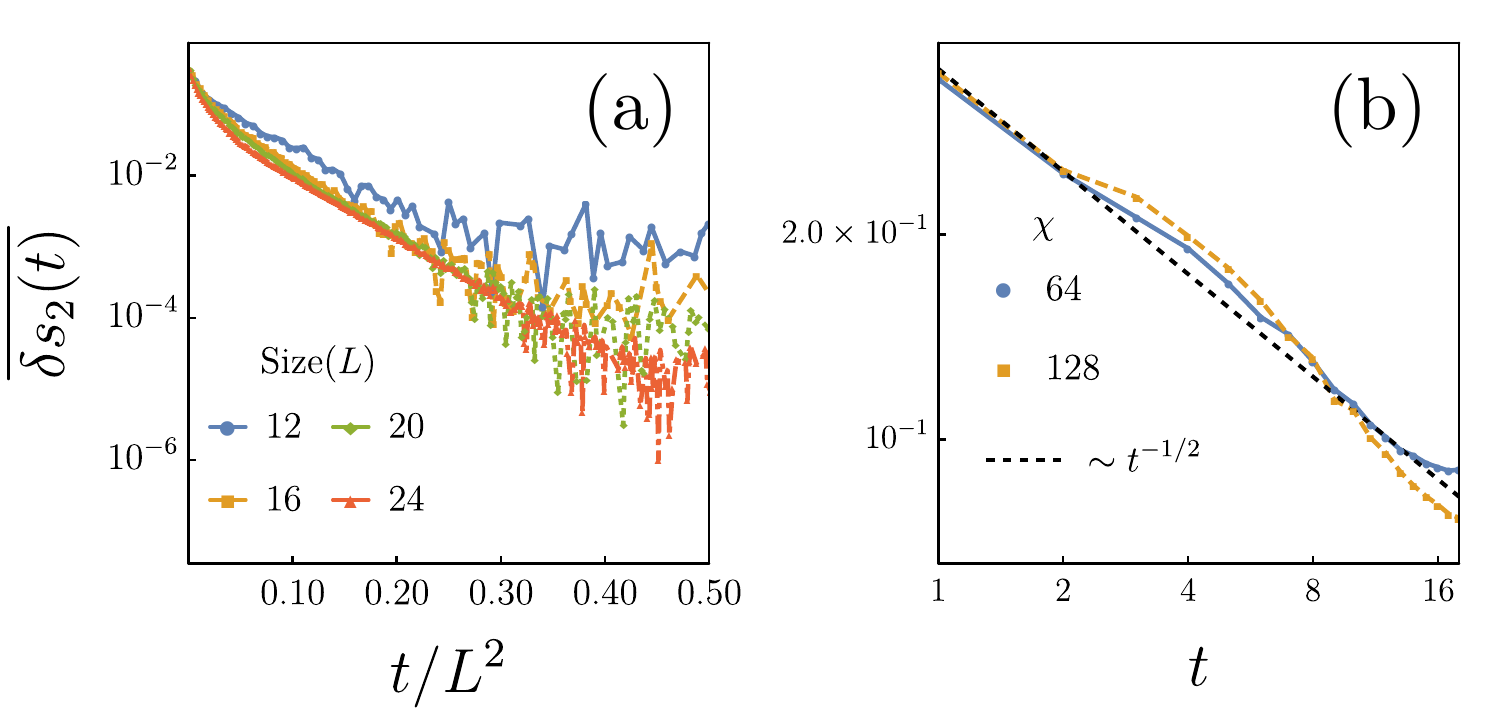}
\caption{%
PE relaxation in the Haar random $\mathrm{U}(1)$ brickwork circuit. Panel (a) shows finite size exact evolution data for the initial state with random configuration with fixed filling factor $\nu = 1/4$ with open boundary condition. Panel (b) shows the infinite chain tensor network calculation with the same filling factor. The initial state is chosen to have a single charge inside each unit cell $L_{\mathrm{cell}} = 1/\nu = 4$. The relaxation follows the same diffusive scaling structure as in the classical Markov model, with nonuniversal quantum amplitudes.}
\label{fig:quantum_entropy_collapse}
\end{figure}

The numerical relaxation is shown in Fig.~\ref{fig:quantum_entropy_collapse}. The finite size exact evolution data exhibit a late time crossover controlled by the diffusive scale $L^2$, while the infinite chain calculation resolves the algebraic relaxation before finite size effects set in. In the same notation as in the classical Markov dynamics, the behavior of the deficit entropy is given by
\begin{equation}
    \overline{\delta s_2(t)}
    \sim
    \begin{cases}
        A_q t^{-1/2}, & 1\ll t\ll L^2,\\
        B_q\exp[-\gamma_q t/L^2], & t\gtrsim L^2,
    \end{cases}
    \label{eq:quantum_entropy_scaling_summary}
\end{equation}
with nonuniversal constants $A_q$, $B_q$, and $\gamma_q$.

Although the near-equilibrium expansion predicts the same hydrodynamic exponent for any fixed R\'enyi index, the quantum simulations presented here focus on $\alpha=2$. Thus, the numerical results test the predicted time dependence at second R\'enyi order and do not imply that the nonuniversal amplitudes are independent of $\alpha$.

\begin{figure}[!htbp]
    \centering
    \includegraphics[width=\columnwidth]{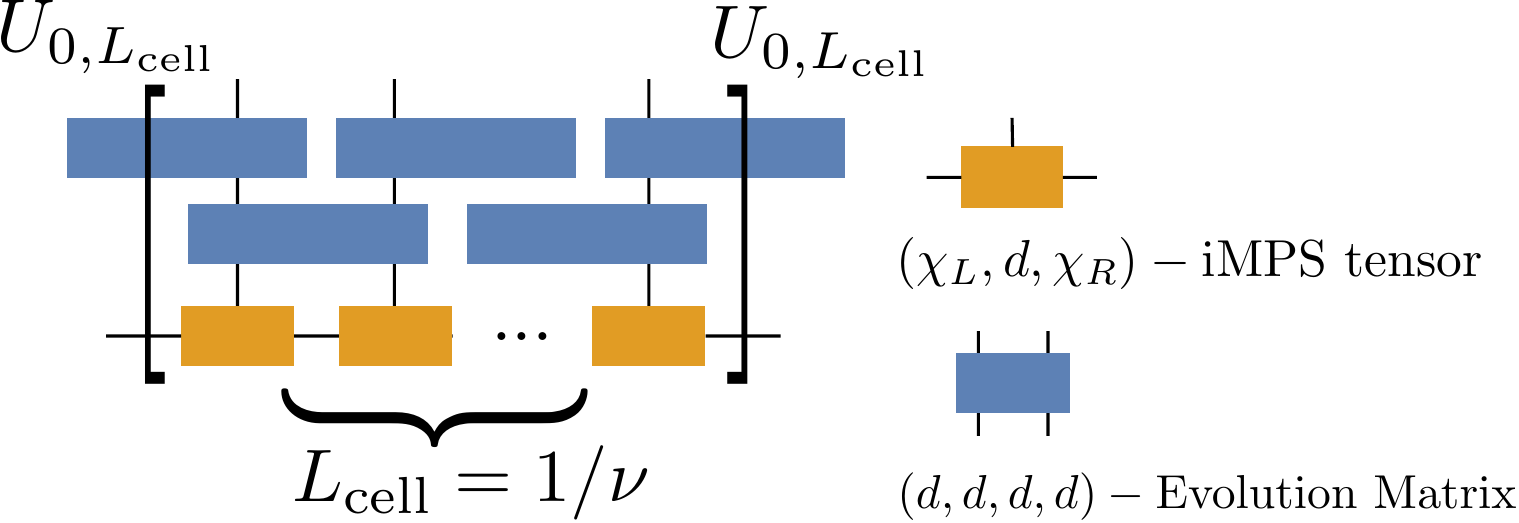}
    \caption{Schematic of the infinite chain iMPS implementation. The tensors and brickwork gates are represented within a finite unit cell that is repeated periodically in space. We take the cell size to be the inverse filling factor $\nu$, $L_{\mathrm{cell}} = 1/\nu$. The gates or evolution matrices inside the unit cell are resampled in time, so the calculation keeps temporal disorder while replacing fully independent spatial disorder by a finite period representative.}
    \label{fig:imps_algorithm}
\end{figure}

We note one technical distinction between the two numerical implementations. In the finite size exact evolution, the Haar random gates are independently drawn in both space and time. In the infinite chain calculation, by contrast, we use a circuit that is random in time but periodic in space. At each time step, we generate a finite spatial unit cell and repeat it periodically along the chain, drawing fresh random $\mathrm{U}(1)$ symmetric gates for each bond within the unit cell, as shown in Fig.~\ref{fig:imps_algorithm}. This construction preserves temporal randomness and the local Haar gate ensemble, while replacing fully independent spatial randomness by finite period spatial randomness. We expect this difference to affect only nonuniversal details, such as amplitudes and transient behavior, and not the hydrodynamic exponent governed by charge diffusion.

Finally, we emphasize a conceptual caveat. The SSEP mapping is exact for the annealed probability distribution $\overline{P_t^\psi(\mathbf{n})}$, whereas Eq.~\eqref{eq:quantum_entropy_deficit} involves a quenched entropy average: the logarithm is taken separately for each circuit realization before averaging. Consequently, the annealed and quenched quantities need not agree in their nonuniversal offsets or amplitudes. Nevertheless, their long time hydrodynamic scaling is controlled by the same conservation law and the same diffusive charge correlations. This is why the quantum PE exhibits the same $t^{-1/2}$ relaxation before the finite size crossover at $t\sim L^2$.

\subsection{Quantum Correlation Diagnostics}

The entropy data alone show the slow relaxation, but the mechanism can be tested more directly by measuring charge correlations. In the quantum iMPS calculation, it is convenient to use the Pauli $Z$ variable
\begin{equation}
    Z_x=2n_x-1 .
\end{equation}
The connected $Z$ correlator is
\begin{equation}
    C_Z(r,t)
    =
    \overline{
    \langle Z_x Z_{x+r}\rangle_t
    -
    \langle Z_x\rangle_t
    \langle Z_{x+r}\rangle_t
    },
    \label{eq:quantum_z_correlator}
\end{equation}
which is related to the connected density correlator by $C_Z(r,t)=4C_n(r,t)$. This normalization changes only the overall prefactor in the squared correlation contribution to Eq.~\eqref{eq:entropy_correlation_decomposition}.

\begin{figure}[!htbp]
    \centering
        \includegraphics[width=\columnwidth]{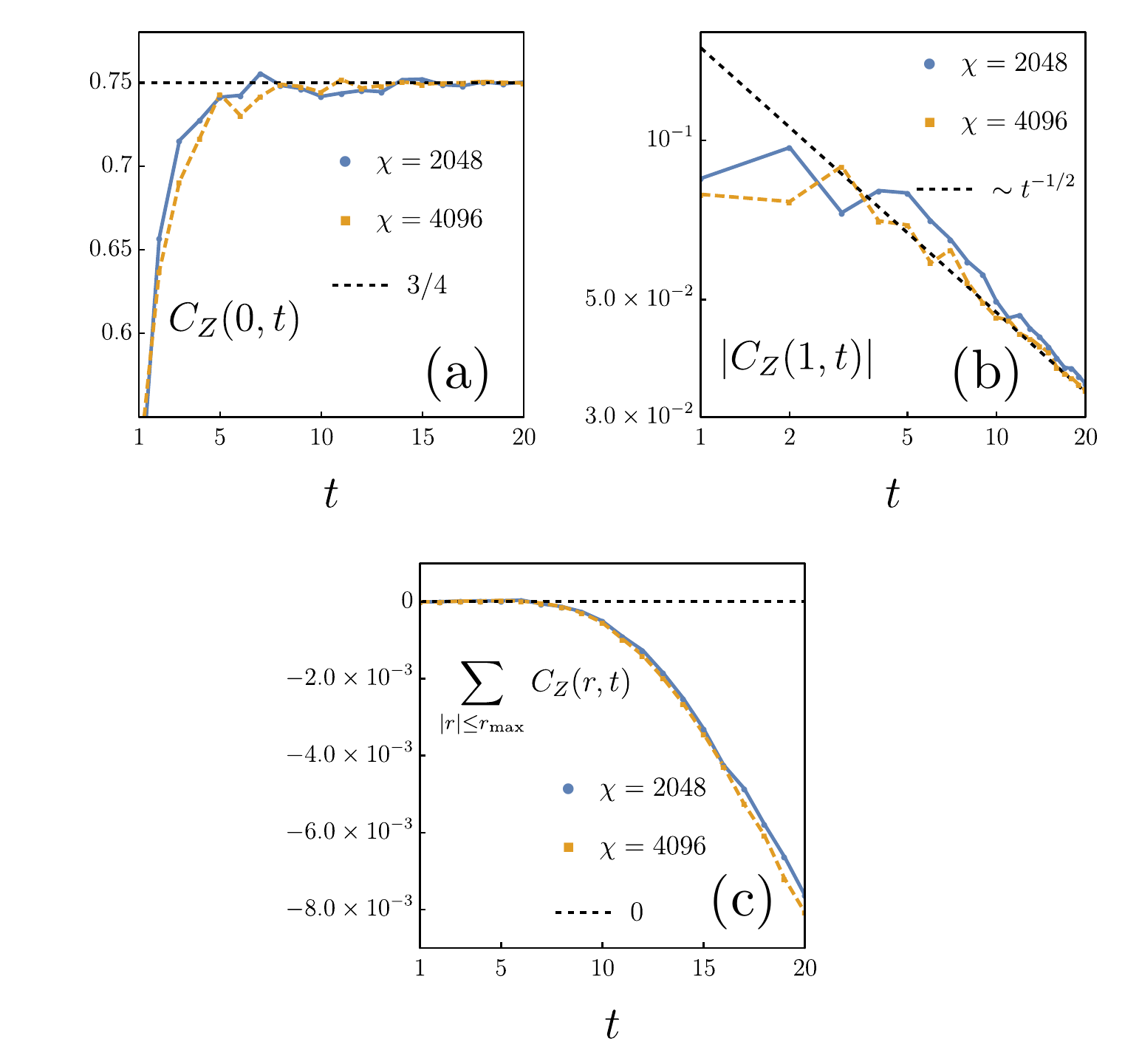}
    \caption{Local and conservation diagnostics for the infinite chain Haar random $\mathrm{U}(1)$ circuit. Panel (a) shows the fast approach of the onsite variance to the local fixed filling value. Panel (b) shows the nearest neighbor absolute connected correlation $|C_Z(1,t)|$ with a $t^{-1/2}$ guide. Panel (c) checks the finite window version of the conservation sum rule in Eq.~\eqref{eq:quantum_correlation_sum_rule}.}
    \label{fig:quantum_corr_local_diagnostics}
\end{figure}

\begin{figure}[!htbp]
    \centering

    \includegraphics[width=\columnwidth]{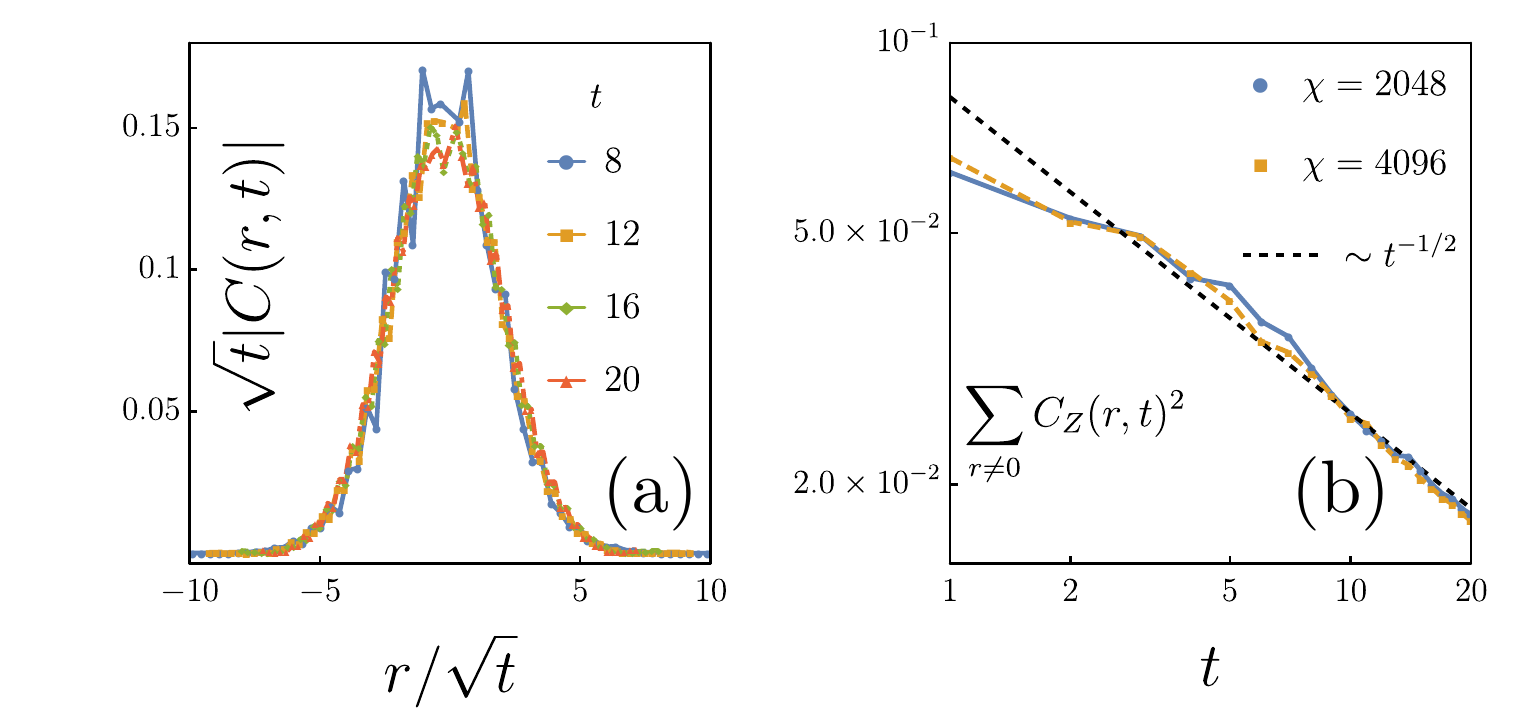}
    \caption{Diffusive scaling diagnostics for the Haar random $U(1)$ circuit. Panel (a) shows the diffusive collapse of the offsite ($r \neq 0$) absolute connected correlation profile as a function of $r/\sqrt{t}$. Panel (b) shows the decay of the offsite squared correlation sum $M_Z(t)$ with a $t^{-1/2}$ guide. }
    \label{fig:quantum_corr_diffusive_scaling}
\end{figure}

At fixed total charge, the same conservation law sum rule applies:
\begin{equation}
    \sum_r C_Z(r,t)=0,
    \qquad
    \sum_{r\neq0}C_Z(r,t)=-C_Z(0,t).
    \label{eq:quantum_correlation_sum_rule}
\end{equation}
Thus the fast local scrambling that sets the onsite variance also creates an opposite sign offsite correlation to compensate, whose long wavelength component spreads diffusively. We therefore expect
\begin{equation}
    C_Z(r,t)
    \simeq
    -t^{-1/2}f_q\!\left(\frac{r}{\sqrt{t}}\right),
    \qquad r\neq0,
    \label{eq:quantum_diffusive_correlation_scaling}
\end{equation}
and consequently
\begin{equation}
    M_Z(t)
    =
    \sum_{r\neq0}C_Z(r,t)^2
    \sim
    t^{-1/2}.
    \label{eq:quantum_corr_squared_sum}
\end{equation}
Figs.~\ref{fig:quantum_corr_local_diagnostics} and \ref{fig:quantum_corr_diffusive_scaling} show these diagnostics. The onsite correlator rapidly approaches the local fixed-filling variance. In the exact calculations we use random product states at fixed filling $\nu=1/4$. In the iMPS simulations we instead start from a period-$L_{\text{cell}}=4$ product state with one particle per unit cell. We work at quarter filling ($\nu=1/4$), rather than the higher filling used in the classical simulations, to improve numerical accuracy in the time window of interest. For the quarter filled Pauli $Z$ data shown, this value is $4\bar{\rho}(1-\bar{\rho})=3/4$. The nearest neighbor correlation develops the expected $t^{-1/2}$ scaling, the finite window sum rule remains approximately satisfied, and the squared offsite sum follows the same $t^{-1/2}$ behavior that appears in the entropy relaxation. The profile collapse confirms that the offsite correlations are organized by the diffusive variable $r/\sqrt{t}$.

\begin{figure}[!htbp]
\centering
\includegraphics[width=\columnwidth]{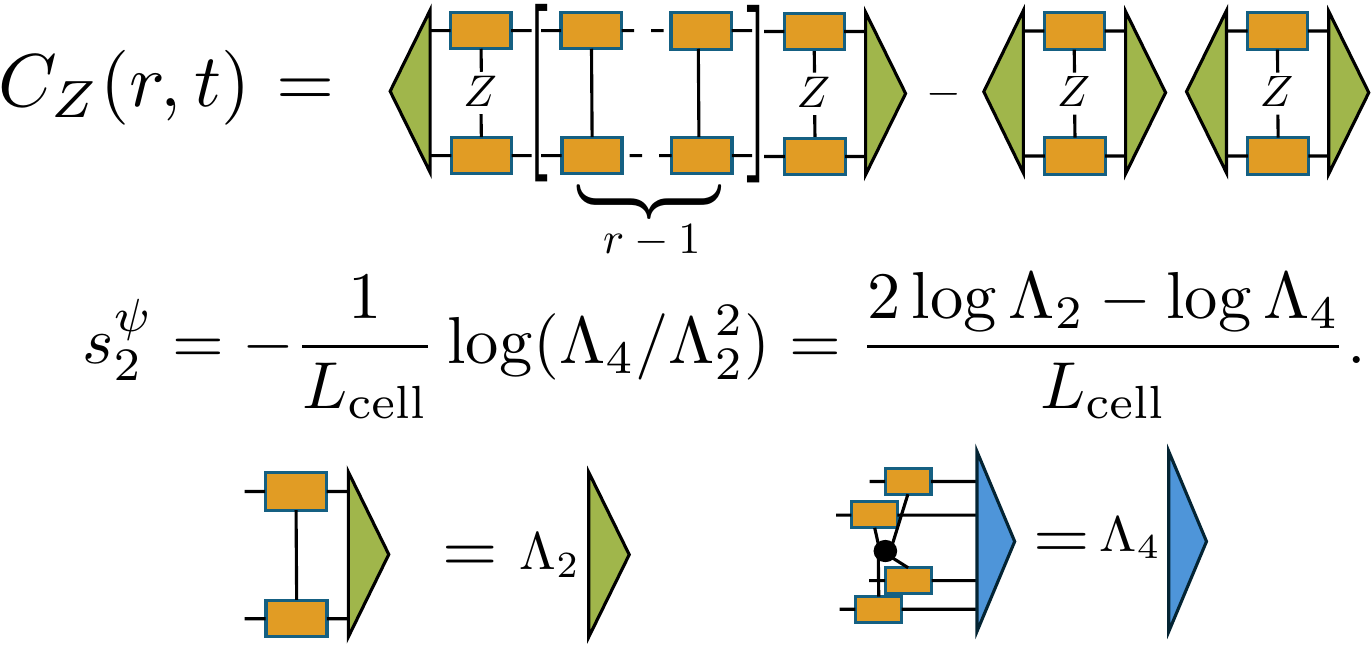}
\caption{Schematic tensor contractions used for the two iMPS observables. The correlation calculation uses single-copy fixed-point environments with local operator insertions, while the PE calculation uses replicated transfer matrices whose dominant eigenvalues are denoted by $\Lambda_2$ and $\Lambda_4$. For clarity, the diagram shows the elementary contraction for a one-site unit cell; for $L_{\mathrm{cell}}>1$, the corresponding transfer matrix is the ordered product of the same local contractions over all tensors in the unit cell, with operator insertions placed on the appropriate sites.}
\label{fig:imps_calculations}
\end{figure}

We close with a technical remark on the computational cost of the two iMPS observables used in this section.  The two iMPS observables investigated in this section, the quantum connected correlation function $C_Z(r,t)$ and the PE density $s_2^\psi$, have very different computational costs, as summarized schematically in Fig.~\ref{fig:imps_calculations}. The TEBD evolution itself is the same in both calculations: each brickwork gate is applied inside the finite unit cell and the enlarged two-site tensor is truncated by an SVD, with leading cost $O(d^3\chi^3)$ for fixed unit-cell size and local dimension $d=2$. Correlations are then evaluated with ordinary single-copy transfer matrices. After computing the left and right iMPS fixed points, one inserts the local operator and propagates the resulting environment through the unit cell to obtain $C_Z(r,t)$ for $|r|\leq l_{\rm max}$; this stores $O(\chi^2)$ environments and involves $O(\chi^3)$ single-copy contractions, with only a linear additional dependence on the number of separations. By contrast, the second PE is a normalized collision-probability observable and therefore requires a replicated Born transfer matrix. Concretely, if $\Lambda_2$ is the dominant eigenvalue of the two-copy norm transfer over one unit cell and $\Lambda_4$ is the dominant eigenvalue of the four-copy Born transfer matrix for $\sum_{\mathbf n}|\psi_{\mathbf n}|^4$, then the iMPS evaluates the PE density as
\begin{equation}
    s_2^\psi
    =
    -\frac{1}{L_{\mathrm{cell}}}
    \log\left(\frac{\Lambda_4}{\Lambda_2^2}\right)
    =
    \frac{2\log\Lambda_2-\log\Lambda_4}{L_{\mathrm{cell}}}.
\label{eq:imps_pe_density_transfer}
\end{equation}
This follows from the standard MPS transfer-matrix contraction of the normalized inverse participation ratio~\cite{Schollwoeck2011MPSReview,Orus2014TensorNetworksIntro}: for $N_{\mathrm{cell}}$ repeated unit cells, $\sum_{\mathbf n}|\psi_{\mathbf n}|^4\sim \Lambda_4^{N_{\mathrm{cell}}}$ and $\langle\psi|\psi\rangle\sim \Lambda_2^{N_{\mathrm{cell}}}$. Thus, besides the normalization transfer with auxiliary dimension $\chi$, one must find the dominant eigenvalue of a fourth-order transfer acting on the squared auxiliary space. In a dense implementation this object has memory cost $O(\chi^4)$ and a site-transfer cost up to $O(d\chi^5)$, and the charge-block implementation reduces prefactors but retains the much steeper dependence on $\chi$. This replicated transfer matrix calculation, rather than the TEBD update, is the numerical bottleneck, which is why the PE data are limited to $\chi=128$ whereas the correlation functions can be pushed to $\chi=4096$.

\section{Quantum $\mathrm{U}(1)$ circuit with five tuning parameters}
\label{sec:quantum_spin_chain}

We finally consider more general dynamics with a conserved $\mathrm{U}(1)$ charge.  The two-site building blocks are the standard charge, or excitation, preserving gates used in XXZ/fSim spin-chain circuits, augmented by local $Z$ phases and a chiral exchange phase~\cite{LiebSchultzMattis1961,Foxen2020FSim,Znidaric2025U1Circuits,Summer2026AnomalousU1Circuits}.  A convenient way to parametrize such charge conserving two-site gates is through local Hermitian generators on neighboring sites $x$ and $x+1$.  Up to an irrelevant identity term, the most general nearest neighbor generator that preserves the two-site charge can be written as
\begin{align}
    h_{x,x+1}
    &=
    a_x Z_x+b_x Z_{x+1}+c_x Z_xZ_{x+1}
    \nonumber\\
    &\quad
    +\frac{\theta_x}{2}X_xX_{x+1}
    +\frac{\theta_x}{2}Y_xY_{x+1}
    \nonumber\\
    &\quad
    +\frac{\phi_x}{2}X_xY_{x+1}
    -\frac{\phi_x}{2}Y_xX_{x+1}.
    \label{eq:u1_spin_chain_hamiltonian}
\end{align}
Here $X_x,Y_x,Z_x$ are Pauli operators and the conserved charge is $Q=\sum_x n_x$, with $n_x=(Z_x+1)/2$ in the convention used above.
The diagonal terms in Eq.~\eqref{eq:u1_spin_chain_hamiltonian} only add phases that depend on the configuration; in spin-chain language these are longitudinal fields and an Ising anisotropy. The $XX+YY$ exchange is the usual $\mathrm{U}(1)$-symmetric XY/XXZ hopping term, while $XY-YX$ is the $z$-directed Dzyaloshinskii-Moriya, or antisymmetric exchange, term~\cite{Dzyaloshinsky1958WeakFerromagnetism,Moriya1960AnisotropicSuperexchange}.  The off-diagonal exchange terms mix $\ket{10}$ and $\ket{01}$ without changing their total charge.  Thus
\begin{equation}
    \left[h_{x,x+1},Q\right]=0,
    \qquad
    \left[\exp(-\mathrm{i}h_{x,x+1}),Q\right]=0 .
    \label{eq:spin_chain_u1_conservation}
\end{equation}

In the brickwork implementation used in the numerics, following common $\mathrm{U}(1)$-symmetric Floquet circuit constructions~\cite{Jonay2024SlowThermalization,Znidaric2025U1Circuits,Summer2026AnomalousU1Circuits}, the two-site unitary is applied as a product of exponentials,
\begin{align}
    U_{x,x+1}
    &=
    \exp\left[-\mathrm{i}
    \left(a_x Z_x+b_x Z_{x+1}+c_x Z_xZ_{x+1}\right)\right]
    \nonumber\\
    &\quad\times
    \exp\left[-\mathrm{i}\frac{\theta_x}{2}
    \left(X_xX_{x+1}+Y_xY_{x+1}\right)\right]
    \nonumber\\
    &\quad\times
    \exp\left[-\mathrm{i}\frac{\phi_x}{2}
    \left(X_xY_{x+1}-Y_xX_{x+1}\right)\right].
    \label{eq:u1_spin_chain_gate}
\end{align}
This is in general different from $\exp(-\mathrm{i}h_{x,x+1})$. The unitary operator for each  time step is
\begin{align}
    U(t)
    &=
    \prod_{x\in \mathrm{even}} U_{x,x+1}(t)
    \prod_{x\in \mathrm{odd}}  U_{x,x+1}(t),
    \label{eq:u1_spin_chain_floquet}
\end{align}
where the parameters may be redrawn in time and from bond to bond. For fixed parameters, this defines a Floquet brickwork unitary, whereas in our simulations we consider a random brickwork circuit. Each factor in Eq.~\eqref{eq:u1_spin_chain_gate} preserves the charge on its bond, so the full brickwork unitary also commutes with $Q$; more generally, locality and Abelian symmetry impose nontrivial constraints on the set of realizable symmetric circuits~\cite{Marvian2024AbelianSymmetryCircuits}. The model has five tuning parameters $a_x$, $b_x$, $c_x$, $\theta_x$, and $\phi_x$, which allow us to control the strength of randomness in the dynamics. When the randomness is too weak, the exact mapping to the SSEP used in Sec.~\ref{sec:quantum_u1} no longer holds. We therefore use this model to test how strongly entropy relaxation is governed by the conserved diffusive charge, and whether the universal behavior predicted in Eq.~\eqref{eq:quantum_entropy_scaling_summary} is observed.

When the dynamics is sufficiently chaotic, we expect the state at long times within a fixed charge sector to have the same local Porter-Thomas fluctuations as the Haar random $\mathrm{U}(1)$ circuit discussed above.  We therefore use the corresponding second PE,
\begin{equation}
    S_{2,\mathrm{Haar}}^{\mathrm{eq}}
    \simeq
    \log\frac{\mathcal{D}+1}{2},
    \qquad
    \mathcal{D}=\binom{L}{N},
    \label{eq:spin_chain_haar_plateau}
\end{equation}
as in Eq.~\eqref{eq:s2_haar_plateau}.

The entropy deficit is again defined as
\begin{equation}
    \overline{\delta s_2(t)}
    =
    \frac{1}{L}(
    S_{2,\mathrm{Haar}}^{\mathrm{eq}}
    -
    \overline{S_2^\psi(t)}).
    \label{eq:spin_chain_entropy_deficit}
\end{equation}

\subsection{Full Random Gates}

The first ensemble is the fully random  gate ensemble.  For each two-site gate application, the five parameters $a_x,b_x,c_x,\theta_x,\phi_x$ are drawn independently from $[0,2\pi)$.

In this random regime, the entropy deficit is expected to follow the same hydrodynamic scaling form as in the Haar random circuit described in Eq.\eqref{eq:quantum_entropy_scaling_summary}.
The fully random data are consistent with this diffusive structure, with ensemble dependent amplitudes and diffusion constant.

\begin{figure}[!htbp]
    \centering
    \includegraphics[width=\columnwidth]{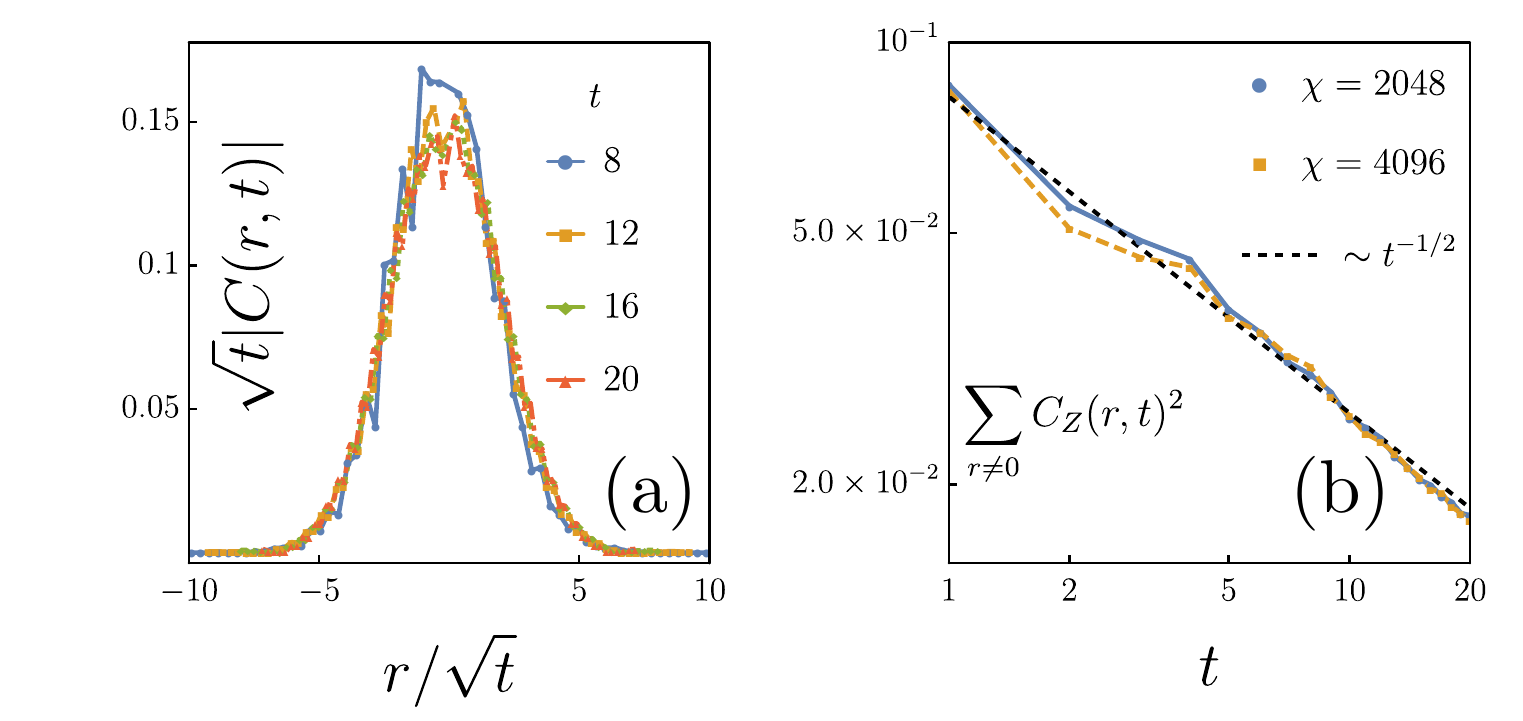}
    \caption{Diffusive correlation diagnostics for the fully random  gate ensemble. Panel (a) shows the collapse of the offsite ($r \neq 0$) connected charge profile as a function of $r/\sqrt{t}$. Panel (b) shows the decay of the offsite squared correlation sum $M_h(t)$ with a $t^{-1/2}$ guide. }
    \label{fig:spin_chain_all_rand_corr_diffusive}
\end{figure}

The correlation diagnostics expose the same mechanism more directly.  After a short microscopic transient, the offsite connected charge correlations spread diffusively,
\begin{equation}
    C_Z(r,t)
    \simeq
    -t^{-1/2}f_h\!\left(\frac{r}{\sqrt{t}}\right),
    \qquad r\neq0,
    \label{eq:spin_chain_all_rand_corr_scaling}
\end{equation}
and the integrated squared weight obeys
\begin{equation}
    M_h(t)
    =
    \sum_{r\neq0} C_Z(r,t)^2
    \sim t^{-1/2}.
    \label{eq:spin_chain_all_rand_corr_squared_sum}
\end{equation}
Fig.~\ref{fig:spin_chain_all_rand_corr_diffusive} shows both the collapse of the rescaled profiles and the $t^{-1/2}$ decay of $M_h(t)$. We use iMPS to simulate dynamics that are random in time but spatially periodic. In the fully random parameter regime, we recover the Haar-random hydrodynamic scaling, although the diffusion constant and overall amplitudes remain ensemble dependent.


\begin{figure}[!htbp]
    \centering

    \includegraphics[width=\columnwidth]{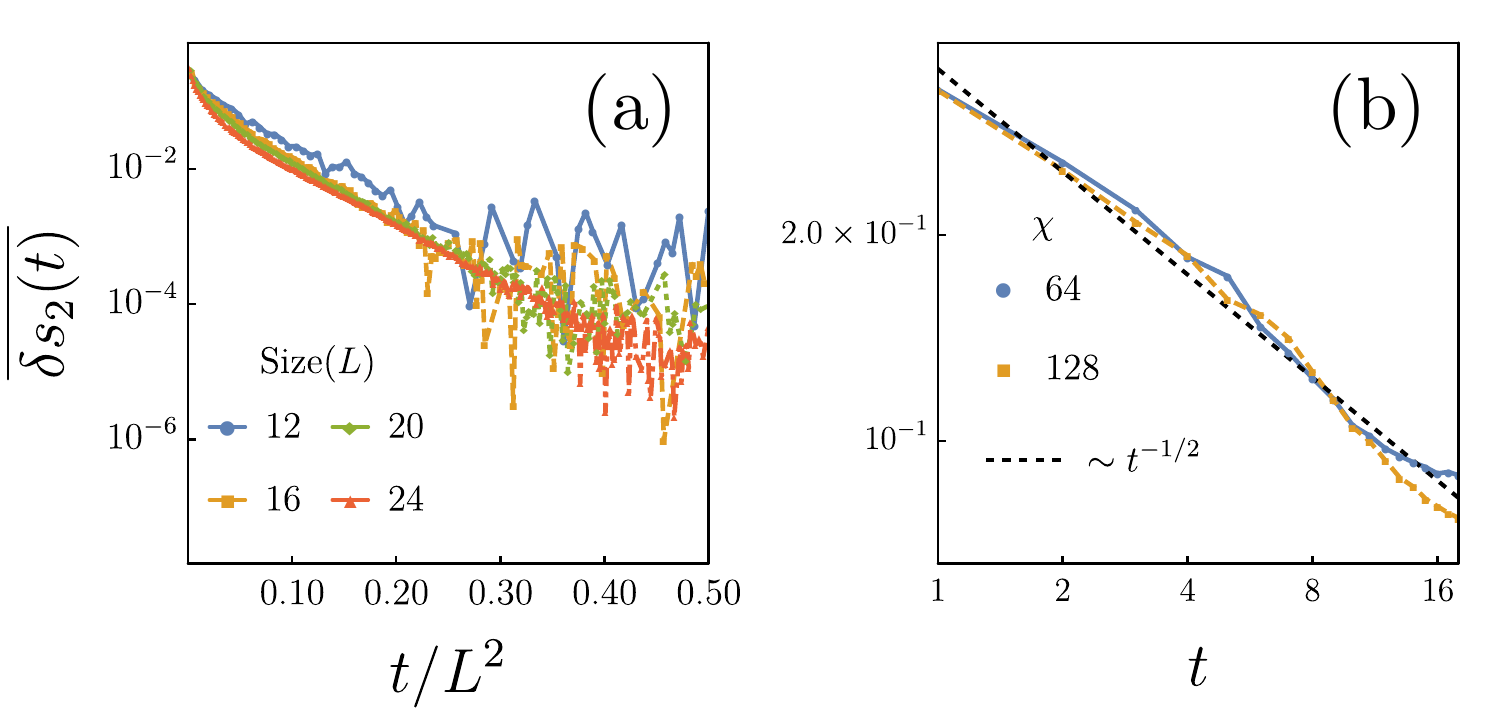}
    \caption{Entropy deficit relaxation for the random circuit with five tuning parameters described in Eq.~\ref{eq:u1_spin_chain_gate}. Panel (a) shows finite size exact evolution data.  The fully random ensemble shows the same diffusive relaxation structure as the Haar random circuit, with ensemble dependent nonuniversal constants. Panel (b) shows the infinite chain calculation, where temporal randomness is retained and the spatial gate pattern is repeated with the iMPS unit cell. }
    \label{fig:spin_chain_all_rand_entropy}
\end{figure}

\subsection{Structured Random Gates}

We also study a more restricted and structured family of $\mathrm{U}(1)$ charge conserving gates with less microscopic randomness.  The staggered longitudinal fields and Ising interactions for all $x$ are fixed to $a_x=0.3\pi$, $b_x=-0.3\pi$, and $c_x=0.3\pi$, while the chiral exchange term is held at $\phi_x=0.5\pi$.  We then vary only the strength and randomness of the $XX+YY$ charge exchange coupling,
\begin{equation}
    \theta_x=\lambda u\pi,\quad
    \lambda\in\{0.1,0.3,0.5,0.7,1\},
    \label{eq:structured_spin_chain_exchange_params}
\end{equation}
where $u$ is independently redrawn from $\mathrm{Unif}(0,1)$ for each gate application for each site $x$.  This sweep isolates the role of randomness in the charge exchange channel while keeping the remaining phase structure fixed.  The staggered field, the chiral exchange, the brickwork ordering, and the temporal randomness move the dynamics away from simple fine-tuned limits, so the question is not whether the gate ensemble is Haar random, but whether the late time relaxation is still governed by the same diffusive charge mode.

\begin{figure}[!htbp]
    \centering

    \includegraphics[width=\columnwidth]{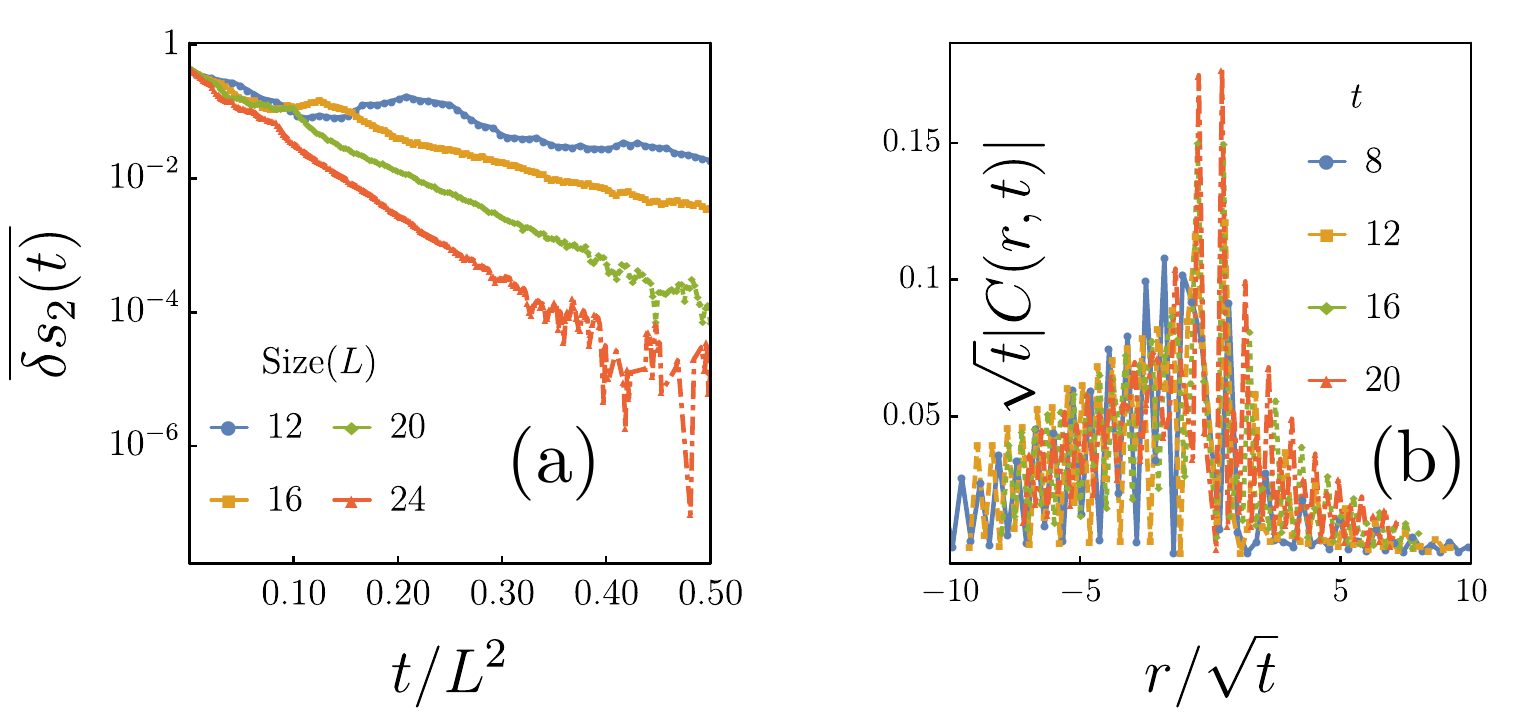}
    \caption{Weak-randomness structured  gate ensemble with $\theta_x=0.1u\pi$ and $u\in\mathrm{Unif}(0,1)$.  Panel (a) shows the entropy deficit relaxation; over the accessible time window, the finite size data do not collapse as a function of $t/L^2$.  Panel (b) shows the offsite ($r \neq 0$) connected charge profile, which likewise does not show a clean diffusive collapse as a function of $r/\sqrt{t}$.}
    \label{fig:spin_chain_01_s2_corr}
\end{figure}

We use the same Porter-Thomas plateau in Eq.~\eqref{eq:spin_chain_haar_plateau} as the entropy reference.  The weakly randomized cases, including $\theta_x=0.1u\pi$, and $\theta_x=0.3u\pi$, show visible nonuniversal deviations from the fully random Hamiltonian gate ensemble over the accessible time window.  Such long transients are natural in weakly randomized charge-conserving Floquet circuits, where proximate integrable or localized regimes can strongly affect finite-time dynamics~\cite{Jonay2024SlowThermalization}.  A representative example with $\theta_x=0.1u\pi$ is shown in Fig.~\ref{fig:spin_chain_01_s2_corr}.  In this weak-randomness regime, the entropy data do not yet exhibit a clean finite size collapse in $t/L^2$, and the offsite charge correlations retain structure that is not organized solely by the diffusive variable $r/\sqrt{t}$.

\begin{figure}[!htbp]
    \centering

    \includegraphics[width=\columnwidth]{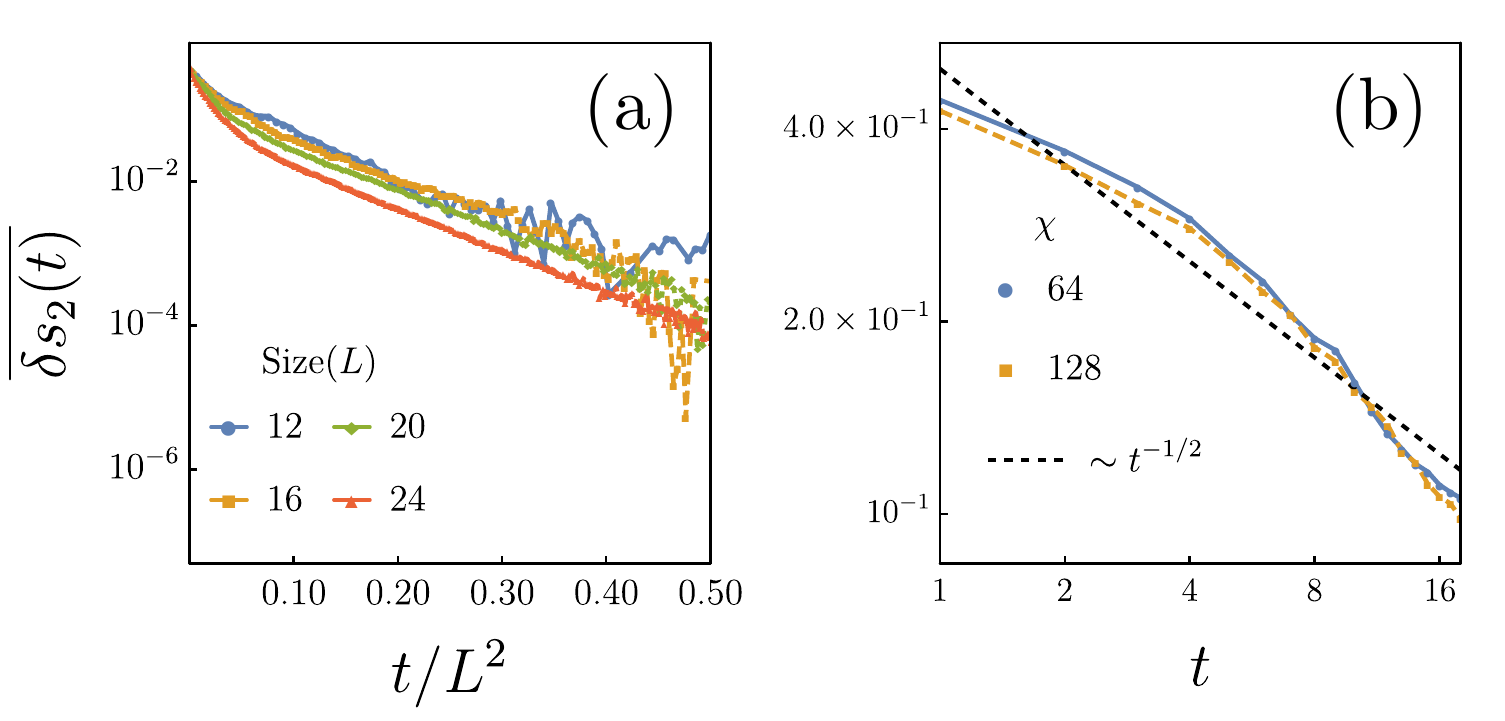}
    \caption{Entropy deficit relaxation for dynamics described in Eq.~\eqref{eq:u1_spin_chain_gate} with $\theta_x=0.7u\pi$ and $u\in\mathrm{Unif}(0,1)$.  Panel (a) shows finite size exact evolution data.  This structured random ensemble shows the same diffusive relaxation structure as the Haar random circuit, with ensemble dependent nonuniversal constants.  Panel (b) shows the infinite chain calculation, where temporal randomness is retained and the spatial gate pattern is repeated with the iMPS unit cell. }
    \label{fig:spin_chain_07_entropy}
\end{figure}

As the exchange randomness is increased, however, the relaxation approaches the same hydrodynamic form.  A representative strong-randomness case is $\lambda=0.7$, shown in Fig.~\ref{fig:spin_chain_07_entropy}, for which
\begin{equation}
    \overline{\delta s_2(t)}
    \sim
    \begin{cases}
        A_s t^{-1/2}, & 1\ll t\ll L^2,\\
        B_s \exp[-\gamma_s t/L^2], & t\gtrsim L^2,
    \end{cases}
    \label{eq:spin_chain_structured}
\end{equation}
with nonuniversal constants $A_s$, $B_s$, and $\gamma_s$ that depend on the structured gate ensemble.

\begin{figure}[!htbp]
    \centering
    \includegraphics[width=\columnwidth]{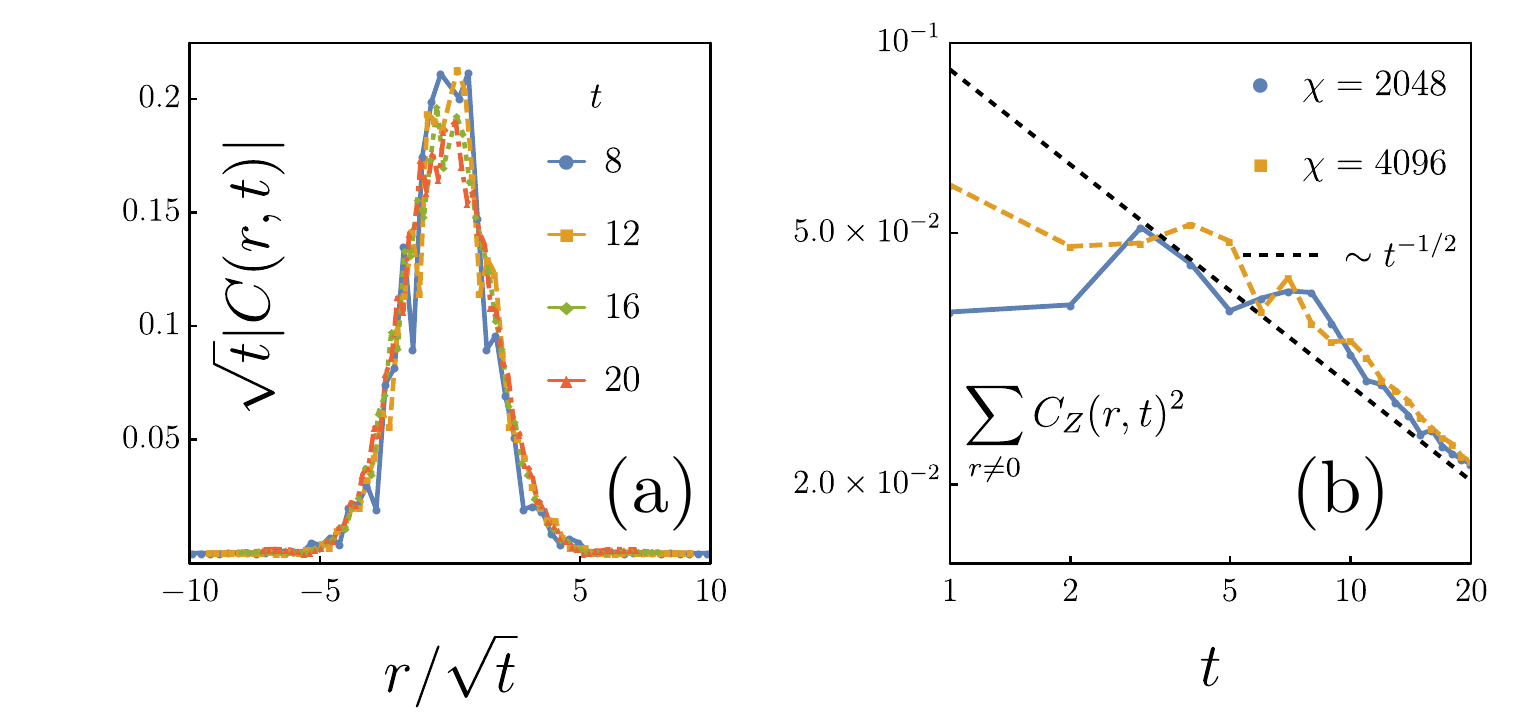}
    \caption{Diffusive correlation diagnostics for the structured Hamiltonian gate ensemble with $\theta_x=0.7u\pi$.    Panel (a) shows the collapse of the offsite connected charge profile as a function of $r/\sqrt{t}$. Panel (b) shows the decay of the offsite ($r\neq 0$) squared correlation sum $M_s(t)$ with a $t^{-1/2}$ guide.}
    \label{fig:spin_chain_07_corr_diffusive}
\end{figure}

The correlation diagnostics show the same crossover more directly.  For the strongly randomized structured gates, the offsite connected charge correlations are compatible with the diffusive scaling form
\begin{equation}
    C_Z(r,t)
    \simeq
    -t^{-1/2}f_s\!\left(\frac{r}{\sqrt{t}}\right),
    \qquad r\neq0,
    \label{eq:spin_chain_structured_corr_scaling}
\end{equation}
and the squared offsite weight obeys
\begin{equation}
    M_s(t)
    =
    \sum_{r\neq0}C_Z(r,t)^2
    \sim
    t^{-1/2}.
    \label{eq:spin_chain_structured_corr_squared_sum}
\end{equation}
Figure~\ref{fig:spin_chain_07_corr_diffusive} shows these diagnostics for $\lambda=0.7$: the rescaled profiles collapse as functions of $r/\sqrt{t}$ and the offsite squared sum follows the same $t^{-1/2}$ envelope that controls the entropy deficit.

Thus the structured ensemble gives a robustness check on the hydrodynamic interpretation.  Once the exchange channel is sufficiently randomized, the PE deficit and the connected charge correlations recover the same diffusive relaxation structure as the Haar random $\mathrm{U}(1)$ circuit and the fully random Hamiltonian gate ensemble.  The microscopic gate distribution changes amplitudes, transient behavior, and the diffusion constant, but it does not change the hydrodynamic exponent in the regime probed by the numerics.  Additional data for the structured random gate sweep are collected in App.~\ref{app:structured_random_gate_sweep}.

\section{Conclusion and outlook}
\label{sec:conclusion}

In this work, we studied the relaxation of PE in charge conserving many-body dynamics. Using a cluster expansion around the equilibrium distribution~\cite{Kubo1962Cumulant,Ruelle1999StatisticalMechanics}, we showed that the entropy deficit at late times is controlled by connected density correlations. More specifically, after local fluctuations and density inhomogeneities have relaxed quickly, the leading contribution to the PE deficit is governed by the square of the connected two-point function. Since charge conservation causes these correlations to relax diffusively~\cite{spohn1991large,kipnis1999scaling}, the PE approaches its equilibrium value on the hydrodynamic time scale, $\tau \sim L^2/D$. This slow relaxation can also be understood from the spectral gap of the corresponding classical master equation. The slowest nonzero mode is a conserved density mode with gap $\Delta_L \sim D/L^2$, while the cluster expansion explains why the PE couples quadratically to these slow modes. Based on this hydrodynamic structure, we proposed a scaling form for the entropy deficit and confirmed it numerically in several classes of random circuits.

Our results suggest that slow PE relaxation should be a generic feature of many-body systems with conservation laws. The precise form of the relaxation can depend on the hydrodynamic universality class, dimensionality, and the structure of the conserved quantities, but the underlying mechanism continues to hold: conservation laws produce long lived, nonlocal density correlations, and the PE is sensitive to these residual correlations. This is consistent with the broader role of conserved diffusive modes in random-circuit dynamics and R\'enyi-type observables~\cite{Rakovszky2018DiffusiveOTOC,Khemani2018OperatorHydro,Rakovszky2019SubballisticRenyi}. Thus, even when conventional local observables appear locally equilibrated, basis resolved quantities such as the PE can retain memory of slow hydrodynamic modes for parametrically long times.

A direct application of this perspective is the relaxation of magic in free fermion random circuits. In such systems, the stabilizer R\'enyi entropy can be expressed as the entropy of a probability distribution over Majorana strings~\cite{Leone2022SRE,Wang2025MonitoredFreeFermionMagic}. Random free fermion unitaries permute Majorana operators within a string while conserving the total number of Majorana operators. Therefore, the induced dynamics of the Majorana string distribution maps naturally onto a classical charge conserving stochastic process, closely related to the SSEP studied here. From this mapping, we expect the relaxation of magic in random free fermion dynamics to exhibit the same slow diffusive behavior as the PE and provides a simple hydrodynamic interpretation of the slow magic relaxation observed numerically in Ref.~\cite{Wang2025MonitoredFreeFermionMagic} \footnote{In that numerical work, the asymptotic diffusive relaxation regime was not resolved because of limitations of the available numerical methods.}.

More broadly, our analysis points to a general connection between magic relaxation and hydrodynamic constraints. For interacting quantum systems with conserved quantities, the distribution over operator strings or stabilizer components may inherit slow modes associated with the underlying conservation laws. We therefore expect analogous slow relaxation of magic in such systems, although the effective stochastic dynamics will generally be more complicated than SSEP. Recent studies of U(1)-constrained nonstabilizerness, coherence, and Hamiltonian/quench dynamics provide useful starting points for this direction~\cite{Tirrito2025UniversalSpreading,Aditya2026CoherenceConservation,Iannotti2026U1Nonstabilizerness,Rattacaso2023StabilizerQuench,odavic2025stabilizer}. For instance, it would be interesting to study magic relaxation in random circuits with charge conservation, as well as in Hamiltonian dynamics with energy conservation.
Developing a systematic theory of magic relaxation in these conserved quantum dynamics, and identifying the associated hydrodynamic universality classes, are interesting directions for future work.

{\it Note}: At the completion of this work, we noted that a recent post arxiv:2604.23192 studied the deficit of quantum coherence in symmetric quantum circuits~\cite{Aditya2026CoherenceConservation}. In what overlaps with this work, they also observed an initial algebraic decay plus an exponential relaxation with diffusive time scale for the PE deficit, albeit the power in the algebraic decay is different. Another recent preprint~\cite{xiao2026diffusivedynamicsnonstabilizerness} reported slow relaxation of quantum magic governed by similar hydrodynamic physics, with particular emphasis on random product states at half filling as initial state. Their power-law exponents for the magic relaxation depend on the R\'enyi index.

\begin{acknowledgements}
   We gratefully acknowledge computing resources from Research Services at Boston College and the assistance provided by Wei Qiu. Gemini 3 and ChatGPT 5.6 Sol were used to assist with parts of the theoretical calculations. 
\end{acknowledgements}

\appendix

\section{Porter-Thomas Reference for the Second PE}
\label{app:pt_plateau}

This appendix derives the Haar random reference value used in Eq.~\eqref{eq:s2_haar_plateau}. This is the standard random-state/Porter-Thomas calculation, using the Gaussian construction of Haar-random vectors and the resulting Dirichlet distribution of intensities~\cite{PorterThomas1956,Mezzadri2007RandomMatrices,ZyczkowskiSommers2001InducedMeasures}.  The calculation is carried out inside a fixed charge sector of Hilbert space dimension
\begin{equation}
    \mathcal{D}=\Omega_N=\binom{L}{N}.
\end{equation}
Let
\begin{equation}
    \ket{\psi}
    =
    \sum_{\alpha=1}^{\mathcal{D}}c_\alpha\ket{\alpha},
    \qquad
    p_\alpha=|c_\alpha|^2,
\end{equation}
where $\ket{\alpha}$ denotes a computational basis state in this sector.  The second PE is
\begin{equation}
    S_2=-\log I_2,
    \qquad
    I_2=\sum_{\alpha=1}^{\mathcal{D}}p_\alpha^2 .
    \label{eq:appendix_s2_ipr}
\end{equation}
The quantity $I_2$ is the inverse participation ratio, or collision probability.

A convenient way to generate a Haar random state is to first draw an unnormalized complex Gaussian vector~\cite{Mezzadri2007RandomMatrices,ZyczkowskiSommers2001InducedMeasures},
\begin{equation}
    g_\alpha=a_\alpha+\mathrm{i} b_\alpha,
\end{equation}
with independent Gaussian real and imaginary parts, and then normalize it:
\begin{equation}
    c_\alpha
    =
    \frac{g_\alpha}
    {\left(\sum_{\beta=1}^{\mathcal{D}}|g_\beta|^2\right)^{1/2}} .
    \label{eq:appendix_gaussian_normalization}
\end{equation}
This produces a Haar random direction because the Gaussian probability density has the form
\begin{equation}
    P(g)\propto
    \exp\left[-\sum_{\alpha=1}^{\mathcal{D}}|g_\alpha|^2\right],
\end{equation}
which depends only on the length of the vector and is invariant under any unitary rotation.  After the normalization in Eq.~\eqref{eq:appendix_gaussian_normalization}, the length has been discarded and only an unbiased random direction remains.

Now define
\begin{equation}
    w_\alpha=|g_\alpha|^2,
    \qquad
    W=\sum_{\beta=1}^{\mathcal{D}}w_\beta .
\end{equation}
For a complex Gaussian variable, $w_\alpha$ is exponentially distributed.  The normalized probabilities are
\begin{equation}
    p_\alpha=\frac{w_\alpha}{W}.
\end{equation}
The vector $(p_1,\ldots,p_{\mathcal{D}})$ is therefore uniformly distributed on the probability simplex
\begin{equation}
    p_\alpha\geq 0,
    \qquad
    \sum_{\alpha=1}^{\mathcal{D}}p_\alpha=1,
\end{equation}
or equivalently has a Dirichlet distribution with all parameters equal to one.  The marginal distribution of one component is
\begin{equation}
    f(p_\alpha)
    =
    (\mathcal{D}-1)(1-p_\alpha)^{\mathcal{D}-2},
    \qquad
    0\leq p_\alpha\leq 1 .
    \label{eq:appendix_beta_distribution}
\end{equation}
For large $\mathcal{D}$ and $p_\alpha=O(1/\mathcal{D})$, this becomes $f(p_\alpha)\simeq \mathcal{D}e^{-\mathcal{D}p_\alpha}$, which is the usual Porter-Thomas distribution for the rescaled intensity $\mathcal{D}p_\alpha$. Its second moment is
\begin{align}
    \overline{p_\alpha^2}
    &=
    (\mathcal{D}-1)
    \int_0^1 dp\, p^2(1-p)^{\mathcal{D}-2}
    \nonumber\\
    &=
    \frac{2}{\mathcal{D}(\mathcal{D}+1)} .
    \label{eq:appendix_p_second_moment}
\end{align}
Summing over the $\mathcal{D}$ basis states gives
\begin{equation}
    \overline{I_2}
    =
    \overline{\sum_{\alpha=1}^{\mathcal{D}}p_\alpha^2}
    =
    \frac{2}{\mathcal{D}+1}.
    \label{eq:haar_ipr}
\end{equation}
For comparison, a perfectly uniform distribution would have $p_\alpha=1/\mathcal{D}$ and hence $I_2=1/\mathcal{D}$.  The Haar random state has a collision probability larger by approximately a factor of two, because the individual probabilities fluctuate around their mean value.  These are the Porter-Thomas fluctuations.

Using Eq.~\eqref{eq:appendix_s2_ipr}, the corresponding second PE is therefore
\begin{equation}
    S_{2,\mathrm{Haar}}^{\mathrm{eq}}
    \simeq
    -\log \overline{I_2}
    =
    \log\frac{\mathcal{D}+1}{2}.
    \label{eq:appendix_s2_pt}
\end{equation}
The symbol $\simeq$ indicates that we use the self averaging Haar value of the inverse participation ratio. The distinction between $-\log \overline{I_2}$ and the trajectory average $\overline{-\log I_2}$ is subleading at large $\mathcal{D}$.

\section{Projection Derivation of the Entropy Correlation Expansion}
\label{app:cluster_projection}

This appendix fills in the steps leading from the second R\'enyi probability space distance in Eq.~\eqref{eq:renyi_alpha_quadratic_distance} to the correlation expansion in Eq.~\eqref{eq:entropy_correlation_decomposition}.  The derivation is a standard orthogonal expansion on a product Bernoulli space, closely related to Fourier analysis on Boolean functions and to cumulant/cluster expansions~\cite{ODonnell2014AnalysisBoolean,Kubo1962Cumulant,Ruelle1999StatisticalMechanics}.  It is simplest if the fixed charge equilibrium measure is approximated locally by an independent Bernoulli measure at density $\bar{\rho}$.  The exact fixed $N$ constraint only changes the zero mode and produces finite size corrections to the orthogonality relations.

Define the equilibrium inner product of two functions of the configuration by
\begin{equation}
    (A,B)_{\mathrm{eq}}
    =
    \sum_{\mathbf{n}}P_{\mathrm{eq}}(\mathbf{n})
    A(\mathbf{n})B(\mathbf{n}) .
    \label{eq:appendix_eq_inner_product}
\end{equation}
Since $P_t=P_{\mathrm{eq}}(1+\epsilon_t)$, any observable $O(\mathbf{n})$ obeys
\begin{align}
    (\epsilon_t,O)_{\mathrm{eq}}
    &=
    \sum_{\mathbf{n}}
    \left[P_t(\mathbf{n})-P_{\mathrm{eq}}(\mathbf{n})\right]
    O(\mathbf{n})
    \nonumber\\
    &=
    \langle O\rangle_t-\langle O\rangle_{\mathrm{eq}} .
    \label{eq:appendix_projection_identity}
\end{align}
This identity is the basic reason why projections of $\epsilon_t$ are given by ordinary density moments.

For the local Bernoulli equilibrium measure,
\begin{equation}
    \langle \delta n_x\rangle_{\mathrm{eq}}=0,
    \qquad
    \langle \delta n_x\delta n_y\rangle_{\mathrm{eq}}
    =
    \kappa\,\delta_{xy},
    \qquad
    \kappa=\bar{\rho}(1-\bar{\rho}) .
    \label{eq:appendix_local_orthogonality}
\end{equation}
Thus distinct products of the fields $\delta n_x$ are orthogonal.  Here $x,y$ label one product and $u,v$ label the product it is being compared with.  For example, for $x<y$ and $u<v$,
\begin{align}
    (\delta n_x,\delta n_u)_{\mathrm{eq}}
    &=
    \kappa\,\delta_{xu},\\
    (\delta n_x,\delta n_u\delta n_v)_{\mathrm{eq}}
    &=
    0,\\
    (\delta n_x\delta n_y,\delta n_u\delta n_v)_{\mathrm{eq}}
    &=
    \kappa^2\,\delta_{xu}\delta_{yv}.
    \label{eq:appendix_product_orthogonality}
\end{align}
The last line is a four point equilibrium average, $\langle\delta n_x\delta n_y\delta n_u\delta n_v\rangle_{\mathrm{eq}}$. For independent local occupations it is nonzero only when the two pairs of sites are the same.  Because each pair is ordered, this gives the single pairing $x=u$, $y=v$, and the value is $\langle\delta n_x^2\rangle_{\mathrm{eq}}\langle\delta n_y^2\rangle_{\mathrm{eq}}=\kappa^2$. Equivalently, if $\phi_A(\mathbf{n})=\prod_{x\in A}\delta n_x$ for a set of sites $A$, then $(\phi_A,\phi_B)_{\mathrm{eq}}=0$ unless $A=B$. These relations are used below as the Gram matrix of the basis: they set the denominators in the projection formula and remove all cross terms when the expansion is inserted into the quadratic entropy distance.

Thus the expansion used in the main text is
\begin{equation}
    \epsilon_t(\mathbf{n})
    =
    \sum_x C_x^{(1)}(t)\delta n_x
    +
    \sum_{x<y}C_{xy}^{(2)}(t)\delta n_x\delta n_y
    +\cdots .
    \label{eq:appendix_cluster_expansion_repeat}
\end{equation}
This is an orthogonal expansion of $\epsilon_t$ in one body terms, two body terms, and higher products of the local density fluctuations.  The coefficient of a basis function is its projection,
\begin{equation}
    C_A(t)
    =
    \frac{(\epsilon_t,\phi_A)_{\mathrm{eq}}}
    {(\phi_A,\phi_A)_{\mathrm{eq}}}.
    \label{eq:appendix_general_projection}
\end{equation}
For a one-site basis function $\phi_{\{x\}}=\delta n_x$, Eqs.~\eqref{eq:appendix_projection_identity} and \eqref{eq:appendix_local_orthogonality} give
\begin{equation}
    C_x^{(1)}(t)
    =
    \frac{(\epsilon_t,\delta n_x)_{\mathrm{eq}}}{\kappa}
    =
    \frac{\langle\delta n_x\rangle_t}{\kappa}.
    \label{eq:appendix_one_body_projection}
\end{equation}
For a two-site basis function $\phi_{\{x,y\}}=\delta n_x\delta n_y$ with $x\neq y$,
\begin{equation}
    C_{xy}^{(2)}(t)
    =
    \frac{(\epsilon_t,\delta n_x\delta n_y)_{\mathrm{eq}}}{\kappa^2}
    =
    \frac{\langle\delta n_x\delta n_y\rangle_t
    -\langle\delta n_x\delta n_y\rangle_{\mathrm{eq}}}
    {\kappa^2}.
    \label{eq:appendix_two_body_projection}
\end{equation}
Within the local product measure the equilibrium term vanishes for $x\neq y$, giving the expression used in the main text.  In the exact fixed $N$ ensemble this term is the small anticorrelation imposed by the global constraint; subtracting it only affects finite size zero mode pieces.

Finally, substitute the orthogonal expansion into the quadratic entropy distance.  Orthogonality removes all cross terms:
\begin{align}
    (\epsilon_t,\epsilon_t)_{\mathrm{eq}}
    &=
    \kappa\sum_x \left[C_x^{(1)}(t)\right]^2
    +
    \kappa^2\sum_{x<y}\left[C_{xy}^{(2)}(t)\right]^2
    +\cdots
    \nonumber\\
    &=
    \frac{1}{\kappa}\sum_x \langle\delta n_x\rangle_t^2
    \nonumber\\
    &\quad+
    \frac{1}{\kappa^2}
    \sum_{x<y}
    \left[
        \langle\delta n_x\delta n_y\rangle_t
        -
        \langle\delta n_x\delta n_y\rangle_{\mathrm{eq}}
    \right]^2
    +\cdots .
    \label{eq:appendix_parseval}
\end{align}
Equation~\eqref{eq:renyi_alpha_quadratic_distance} says $\Delta S_\alpha=\frac{\alpha}{2}(\epsilon_t,\epsilon_t)_{\mathrm{eq}}+\cdots$, which gives Eq.~\eqref{eq:entropy_correlation_decomposition}.  To connect the two-site moment to the connected correlator, write
\begin{equation}
    \langle\delta n_x\delta n_y\rangle_t
    =
    C(x,y;t)
    +
    \langle\delta n_x\rangle_t
    \langle\delta n_y\rangle_t .
    \label{eq:appendix_moment_connected_relation}
\end{equation}
For the N\'eel quench, the one-point terms decay exponentially after the fast staggered mode relaxes.  In that late time regime the two body part of the entropy gap is therefore controlled by the squared connected correlator, as used in Eq.~\eqref{eq:entropy_correlation_decomposition}.

\section{Finite size crossover from the diffusive correlation mode}
\label{app: crossover}

In the main text we used the standard diffusive scaling argument for conserved-density correlations~\cite{spohn1991large,kipnis1999scaling},
\begin{equation}
    C(r,t) \sim t^{-1/2} f\left(\frac{r}{\sqrt{t}}\right)
\end{equation}
to explain the algebraic entropy relaxation in the thermodynamic diffusive window $1\ll t\ll L^2$. Here we show how the same entropy correlation relation gives the finite size late time form for $t\gtrsim L^2$.

After the fast one body density mode has decayed, the leading contribution to the second PE deficit comes from the connected two-point sector,
\begin{equation}
    \Delta S_2(t)
    \simeq
    \frac{1}{\kappa^2}
    \sum_{x<y} C(x,y;t)^2 ,
    \label{eq:appendix_s2_corr}
\end{equation}
where $C(x,y;t)=\langle n_x n_y\rangle_t-\langle n_x\rangle_t\langle n_y\rangle_t$. For a translation invariant periodic chain, $C(x,y;t)=C(r,t)$, with $r=y-x$, so
\begin{equation}
    \delta s_2(t)
    =
    \frac{\Delta S_2(t)}{L}
    \sim
    \frac{1}{\kappa^2}\sum_r C(r,t)^2 ,
    \label{eq:appendix_s2_density_corr}
\end{equation}
up to an unimportant numerical factor.

The long wavelength evolution of the connected correlator is diffusive. On a finite periodic chain this means
\begin{equation}
    \partial_t C(r,t)=D\partial_r^2 C(r,t),
    \label{eq:appendix_diff_eq}
\end{equation}
with periodic boundary conditions. Expanding in Fourier modes,
\begin{equation}
    C(r,t)=\sum_m c_m(t)e^{ik_m r},
    \qquad
    k_m=\frac{2\pi m}{L},
\end{equation}
gives
\begin{equation}
    c_m(t)=c_m(0)e^{-Dk_m^2t}.
\end{equation}
The zero mode is absent because the fixed charge constraint imposes the sum rule
\begin{equation}
    \sum_r C(r,t)=0.
\end{equation}
Therefore the slowest surviving mode has $k_1=2\pi/L$. At late times,
\begin{equation}
    C(r,t)
    \simeq
    A\cos\left(\frac{2\pi r}{L}+\phi\right)
    \exp\left[-D\left(\frac{2\pi}{L}\right)^2 t\right],
    \label{eq:appendix_slowest_mode}
\end{equation}
where the amplitude $A$ and phase $\phi$ are nonuniversal.

Substituting Eq.~\eqref{eq:appendix_slowest_mode} into Eq.~\eqref{eq:appendix_s2_density_corr}, we find
\begin{align}
    \delta s_2(t)
    &\sim
    \frac{A^2}{\kappa^2}
    \exp\left[-2D\left(\frac{2\pi}{L}\right)^2t\right]
    \sum_r
    \cos^2\left(\frac{2\pi r}{L}+\phi\right) .
\end{align}
The spatial sum only contributes an $O(L)$ factor, which is cancelled by the normalization to the entropy density. Thus the late time decay is
\begin{equation}
    \delta s_2(t)
    \sim
    B\exp\left[-\gamma\frac{t}{L^2}\right],
    \qquad
    \gamma = 8\pi^2 D ,
    \label{eq:appendix_late_time_exp}
\end{equation}
up to convention dependent factors in the definition of one time step and of the diffusion constant. The factor of two in the exponent compared with the decay rate of the slowest correlator mode comes from the fact that the entropy deficit is quadratic in $C(r,t)$. This gives the finite size late time form quoted in Eq.~\eqref{eq:markov_entropy_scaling_summary}.

\section{Additional Structured Random Gate Data}
\label{app:structured_random_gate_sweep}

\newcommand{\appstructuredpanelwidth}{0.47\columnwidth}

\begin{figure}[t]
    \centering
    \subfloat[$\lambda=0.1$.\label{fig:app_structured_ed_entropy_01}]{
        \includegraphics[width=\appstructuredpanelwidth]{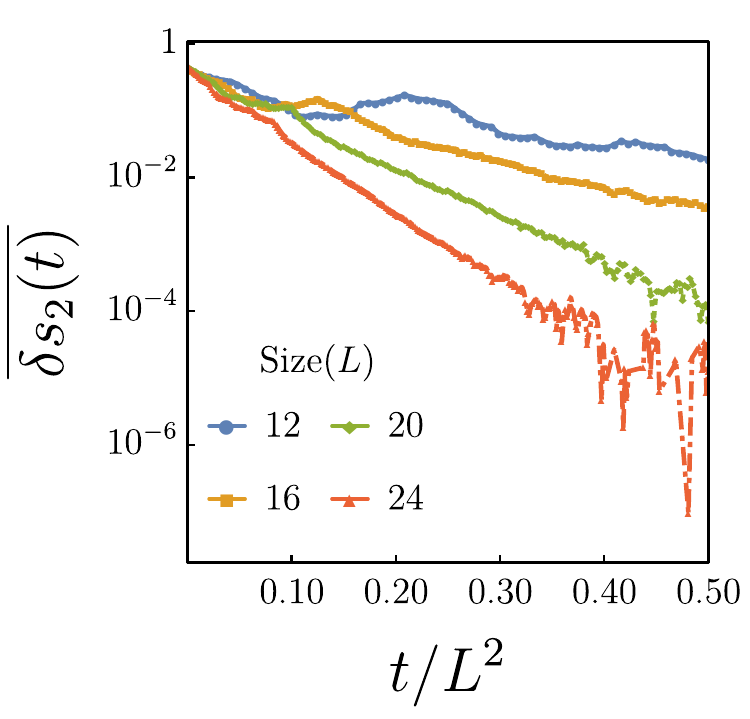}}
    \hfill
    \subfloat[$\lambda=0.3$.\label{fig:app_structured_ed_entropy_03}]{
        \includegraphics[width=\appstructuredpanelwidth]{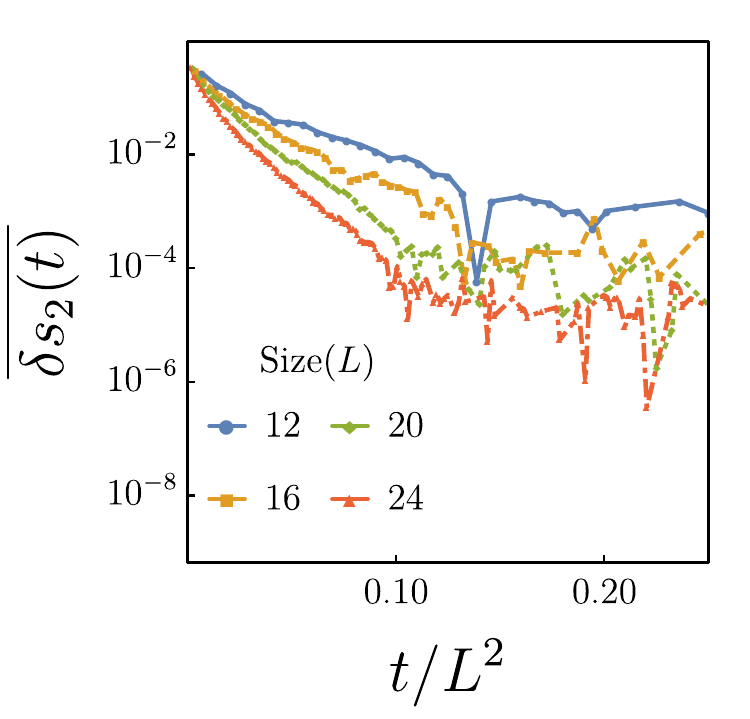}}
    \\[0.1em]
    \subfloat[$\lambda=0.5$.\label{fig:app_structured_ed_entropy_05}]{
        \includegraphics[width=\appstructuredpanelwidth]{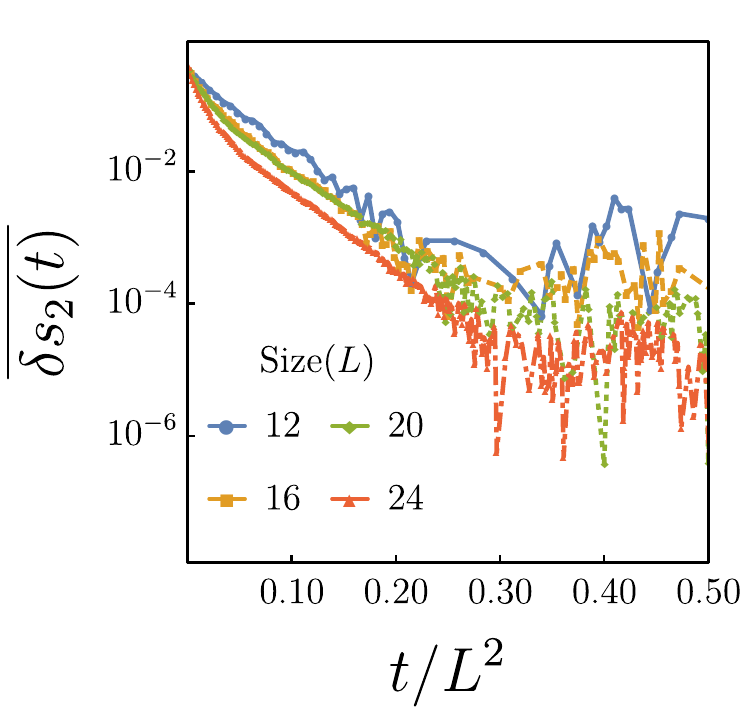}}
    \hfill
    \subfloat[$\lambda=0.7$.\label{fig:app_structured_ed_entropy_07}]{
        \includegraphics[width=\appstructuredpanelwidth]{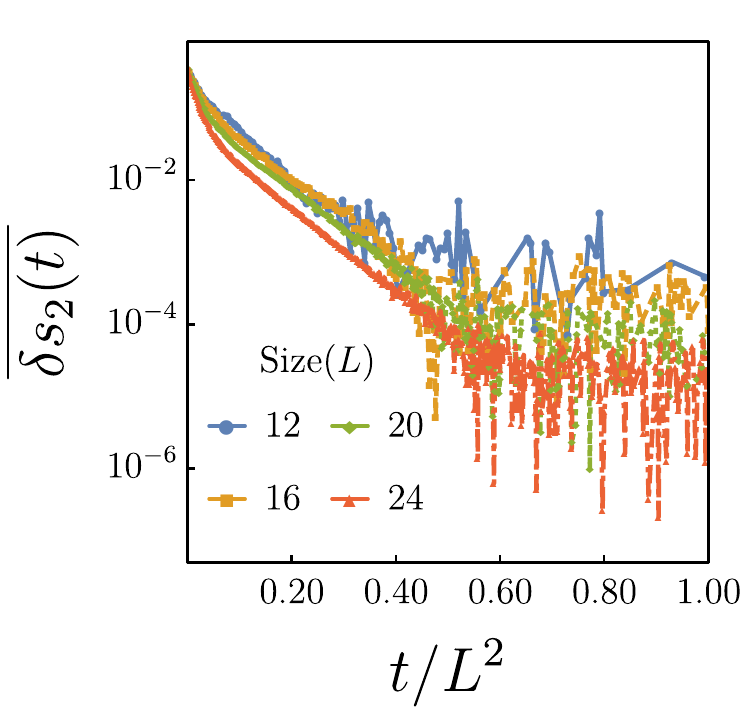}}
    \\[0.1em]
    \subfloat[$\lambda=1$.\label{fig:app_structured_ed_entropy_1}]{
        \includegraphics[width=\appstructuredpanelwidth]{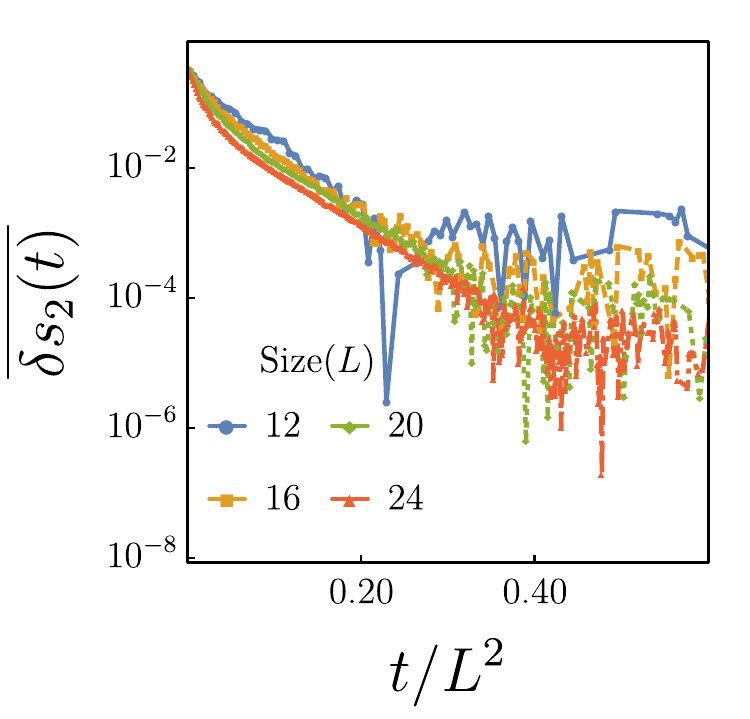}}
    \caption{Finite size exact evolution entropy-deficit collapse for the structured random gate ensemble with $\theta_x=\lambda u\pi$, plotted as $\overline{\delta s_2(t)}$ against $t/L^2$.}
    \label{fig:app_structured_entropy_ed_sweep}
\end{figure}

\begin{figure}[t]
    \centering
    \subfloat[$\lambda=0.1$.\label{fig:app_structured_imps_entropy_01}]{
        \includegraphics[width=\appstructuredpanelwidth]{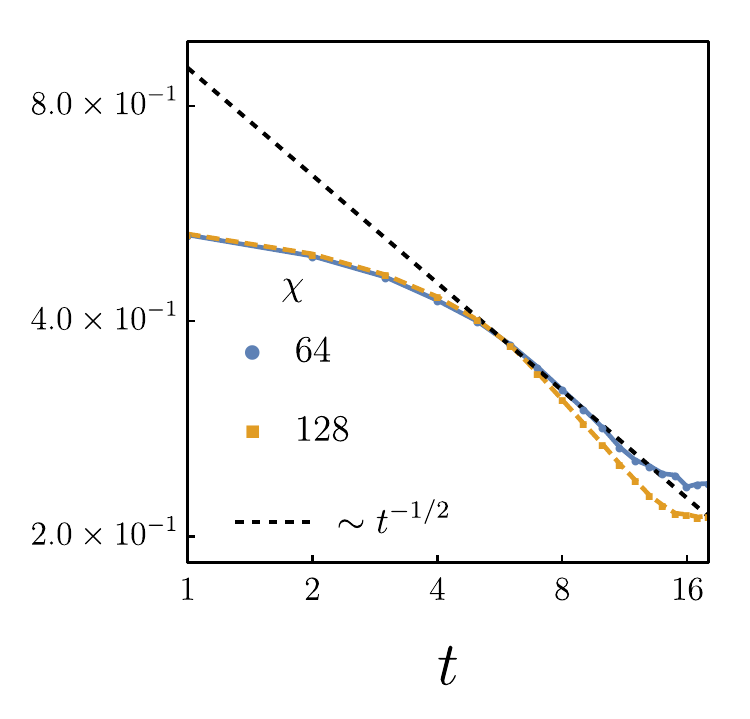}}
    \hfill
    \subfloat[$\lambda=0.3$.\label{fig:app_structured_imps_entropy_03}]{
        \includegraphics[width=\appstructuredpanelwidth]{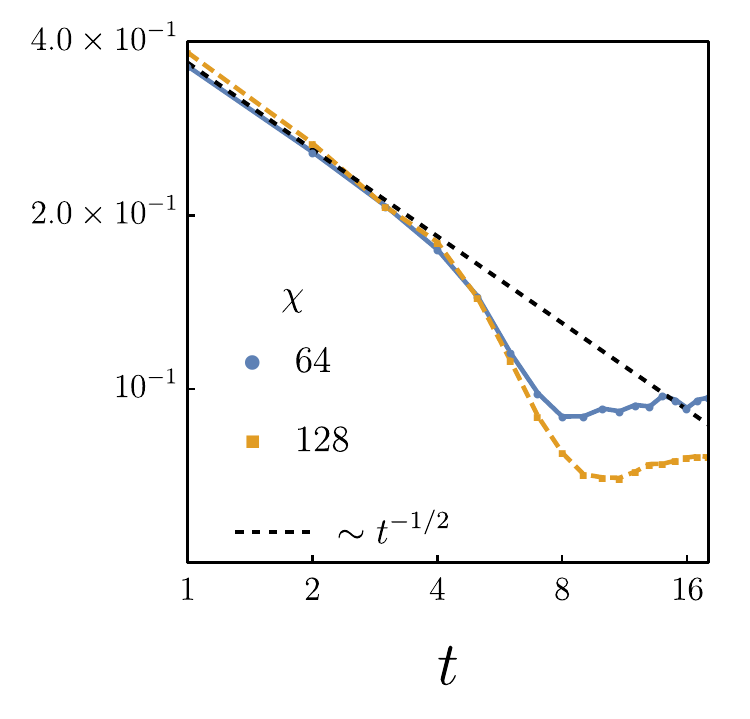}}
    \\[0.1em]
    \subfloat[$\lambda=0.5$.\label{fig:app_structured_imps_entropy_05}]{
        \includegraphics[width=\appstructuredpanelwidth]{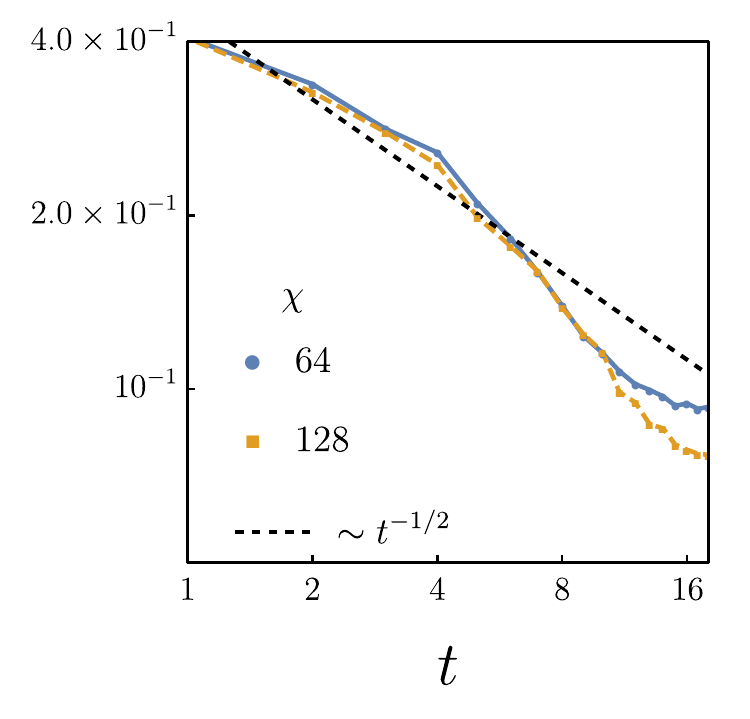}}
    \hfill
    \subfloat[$\lambda=0.7$.\label{fig:app_structured_imps_entropy_07}]{
        \includegraphics[width=\appstructuredpanelwidth]{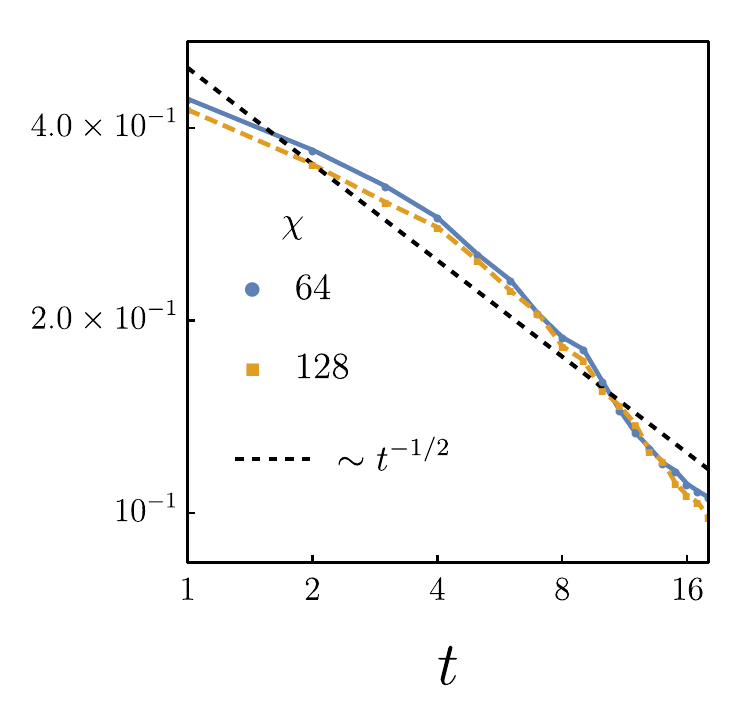}}
    \\[0.1em]
    \subfloat[$\lambda=1$.\label{fig:app_structured_imps_entropy_1}]{
        \includegraphics[width=\appstructuredpanelwidth]{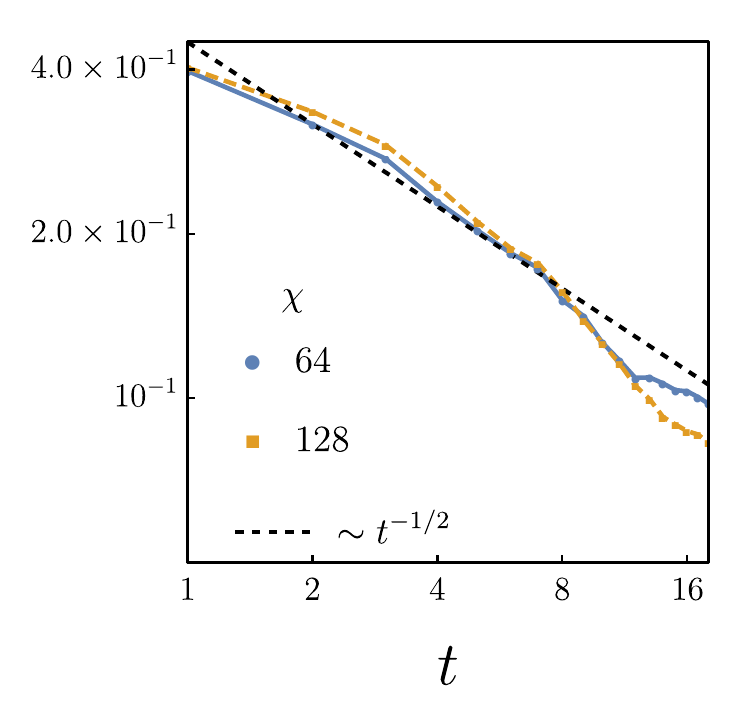}}
    \caption{Infinite chain iMPS entropy-deficit relaxation for the structured random gate ensemble with $\theta_x=\lambda u\pi$.}
    \label{fig:app_structured_entropy_imps_sweep}
\end{figure}

\begin{figure}[t]
    \centering
    \subfloat[$\lambda=0.1$.\label{fig:app_structured_corr_01}]{
        \includegraphics[width=\appstructuredpanelwidth]{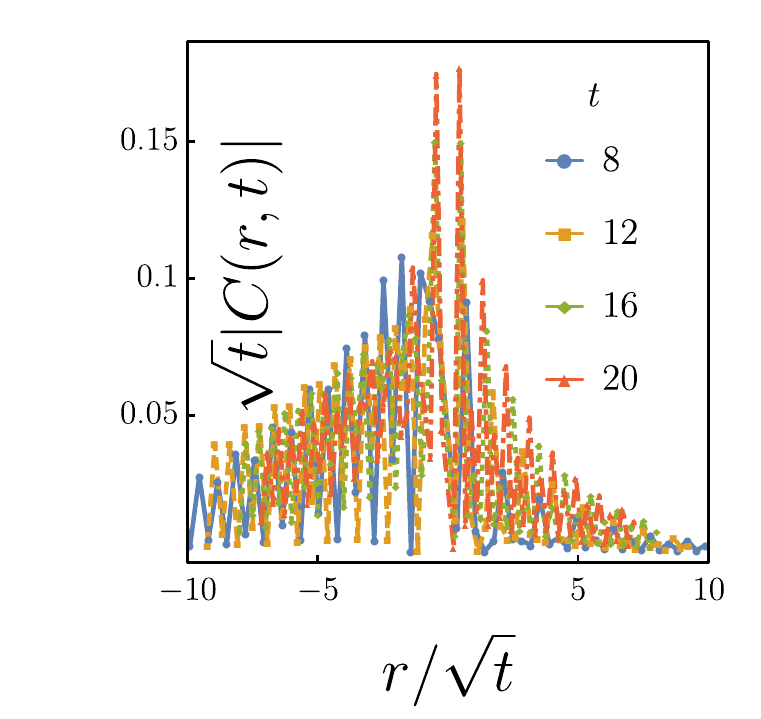}}
    \hfill
    \subfloat[$\lambda=0.3$.\label{fig:app_structured_corr_03}]{
        \includegraphics[width=\appstructuredpanelwidth]{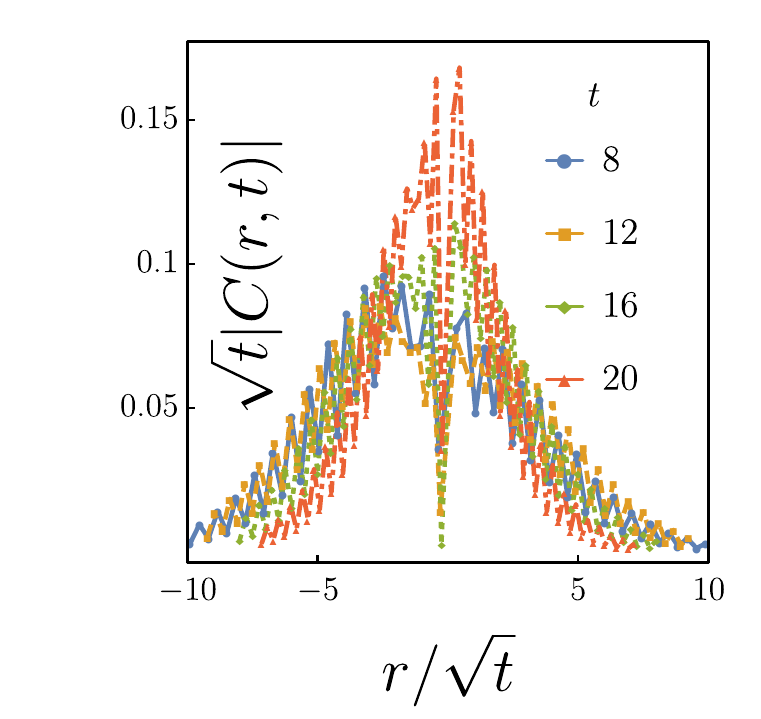}}
    \\[0.1em]
    \subfloat[$\lambda=0.5$.\label{fig:app_structured_corr_05}]{
        \includegraphics[width=\appstructuredpanelwidth]{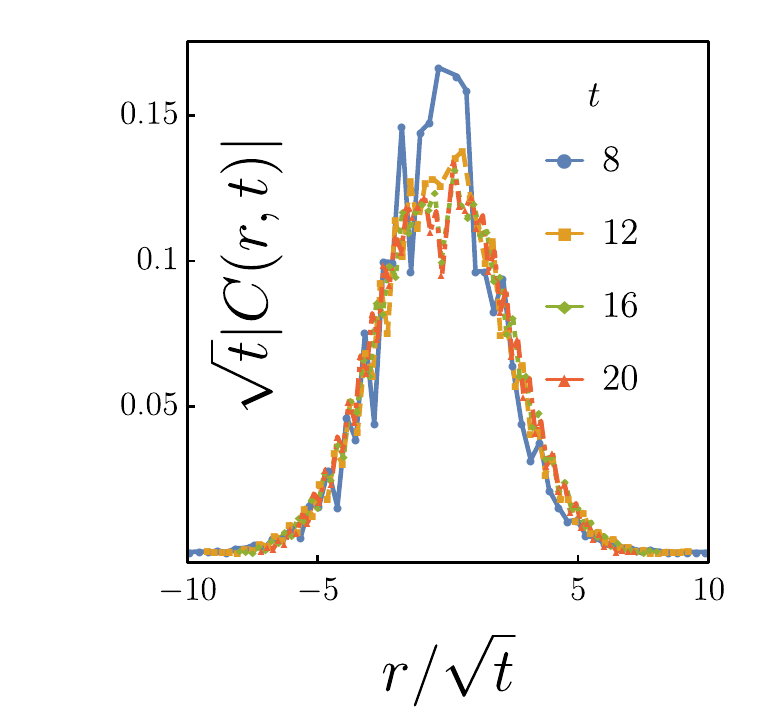}}
    \hfill
    \subfloat[$\lambda=0.7$.\label{fig:app_structured_corr_07}]{
        \includegraphics[width=\appstructuredpanelwidth]{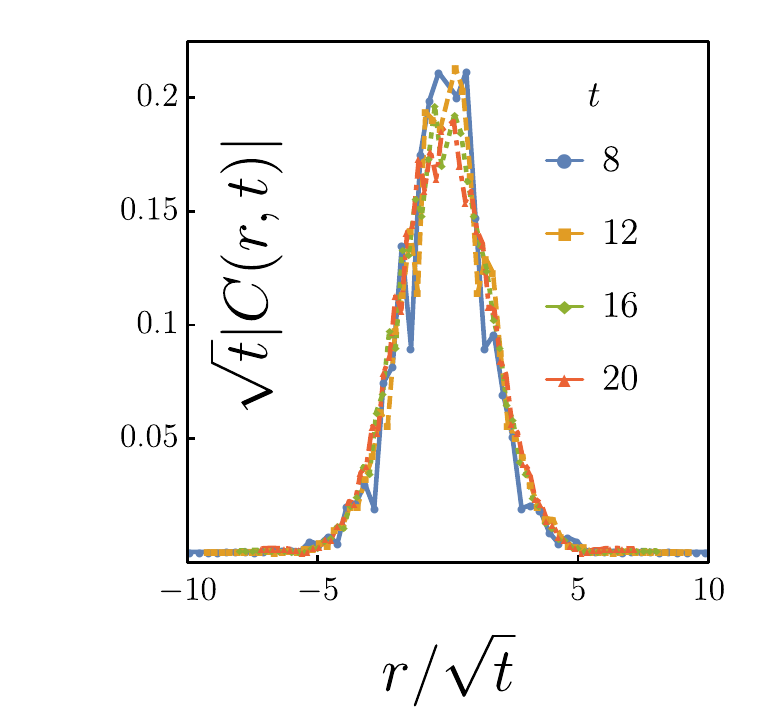}}
    \\[0.1em]
    \subfloat[$\lambda=1$.\label{fig:app_structured_corr_1}]{
        \includegraphics[width=\appstructuredpanelwidth]{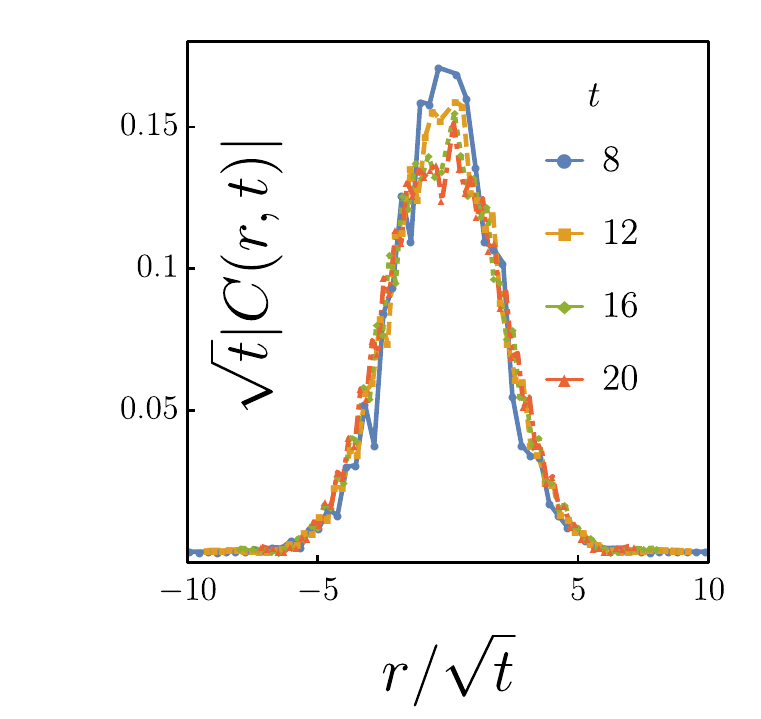}}
    \caption{Diffusive collapse test for the connected charge correlations in the structured random gate ensemble with $\theta=\lambda u\pi$.  Each panel shows the offsite ($r \neq 0$) profile, rescaled according to the diffusive form $C_Z(r,t)\simeq -t^{-1/2}f_s(r/\sqrt{t})$, as a function of $r/\sqrt{t}$.}
    \label{fig:app_structured_corr_collapse_sweep}
\end{figure}

This appendix collects supplementary numerical data for the structured random gate family introduced in Eq.~\eqref{eq:structured_spin_chain_exchange_params}.  The parameters $a_x=0.3\pi$, $b_x=-0.3\pi$, $c_x=0.3\pi$, and $\phi_x=0.5\pi$ are held fixed for all $x$, while the exchange angle is drawn as $\theta_x=\lambda u\pi$ with $u\in\mathrm{Unif}(0,1)$ for each $x$.  The main text shows representative weak- and strong-randomness cases.  Here we summarize the sweep over $\lambda$ to show how the entropy relaxation and charge-correlation profiles cross over toward the diffusive scaling structure.

Figs~\ref{fig:app_structured_entropy_ed_sweep} and \ref{fig:app_structured_entropy_imps_sweep} show the entropy deficit data.  The finite size exact evolution data test the late time scaling collapse controlled by the diffusive variable $t/L^2$, while the infinite chain iMPS data probe the thermodynamic relaxation window before finite size effects set in.  In both calculations, increasing $\lambda$ strengthens the random charge exchange channel and reduces the nonuniversal transient structure.

The correlation diagnostics in Fig.~\ref{fig:app_structured_corr_collapse_sweep} test whether the offsite connected charge profiles are organized by the diffusive scaling variable $r/\sqrt{t}$.  For weak exchange randomness the profiles retain visible microscopic structure over the simulated time window, while stronger randomization gives cleaner collapse.

\bibliography{ref}

\end{document}